\documentclass[12pt]{article}

\ifx\pdfoutput\undefined
\usepackage[dvips,bookmarks]{hyperref}
\else
\usepackage{hyperref}
\fi
\hypersetup{colorlinks=false,bookmarksopen,bookmarksnumbered,citecolor=blue,
   pdfstartview=FitH}

\usepackage{latexsym}

\usepackage{amssymb,amsfonts,amsmath}
\usepackage{graphicx}
\usepackage{indentfirst}
\usepackage{enumitem}
\usepackage{bbm}
\usepackage{cite}

\parskip=\medskipamount

\newcommand {\cD}{{\cal D}}
\newcommand {\cE}{{\cal E}}

\newcommand {\cN}{{\cal N}}

\def\a{\alpha}

\def\b{\beta}

\def\d{\delta}

\def\f{\phi}
\def\g{\gamma}
\def\G{\Gamma}

\def\j{\psi}
\def\k{\kappa}
\def\l{\lambda}
\def\m{\mu}

\def\o{\omega}
\def\p{\pi}
\def\q{\theta}
\def\r{\rho}
\def\s{\sigma}

\def\z{\zeta}
\def\D{\Delta}
\def\F{\Phi}
\def\J{\Psi}
\def\L{\Lambda}
\def\O{\Omega}

\def\X{\Xi}
\def\tr{{\rm tr}}
\def\Tr{{\rm Tr}}
\def\rd{{\rm d}}
\def\ri{{\rm i}}
\def\re{{\rm e}}

\def\h{\eta}
\def\w{\omega}
\def\ce{\varepsilon}
\newcommand{\ad}{{\dot{\alpha}}}                           
\newcommand{\cDB}{{\bar\cD}}                            
\newcommand{\DB}{\bar{D}}

\newcommand{\pd}{\partial}                           
\newcommand{\hf}{\frac12}

\newcommand{\vf}{\varphi}
\newcommand{\pa}{\partial}
\counterwithin*{equation}{section}

\newcommand{\be}{\begin{equation}}
\newcommand{\ee}{\end{equation}}
\newcommand{\bea}{\begin{eqnarray}}
\newcommand{\eea}{\end{eqnarray}}
\newcommand{\non}{\nonumber}

\newcommand{\dsR}{{\mathbb R}}

\newcommand{\bm}[1]{\mbox{\boldmath$#1$}}

\def\double #1{#1{\hbox{\kern-2pt $#1$}}}

\newcommand{\no}{\nonumber}
\newcommand{\ba}{\bar}
\newcommand{\ii}{\mathrm{i}}
\newcommand{\ex}{\mathrm{e}}
\newcommand{\dd}{\mathrm{d}}

\newcommand{\sSp}{\mathsf{Sp}}
\newcommand{\sSU}{\mathsf{SU}}
\newcommand{\sSL}{\mathsf{SL}}

\newcommand{\sSO}{\mathsf{SO}}
\newcommand{\sU}{\mathsf{U}}

\newcommand{\sMat}{\mathsf{Mat}}

\newcommand{\bsubeq}{\begin{subequations}}
\newcommand{\esubeq}{\end{subequations}}

\def\rT{{\rm T}}

\newcommand{\op}{\Delta}
\renewcommand{\~}{\tilde}
\renewcommand{\.}{\dot}
\newcommand{\ph}{\phantom}
\begin{document}

\begin{titlepage}
\begin{flushright}
February, 2023\\
\end{flushright}
\vspace{5mm}

\begin{center}
{\Large \bf Effective actions in supersymmetric gauge theories:
heat kernels for non-minimal operators
 }
\end{center}

\begin{center}

{\bf

Darren T. Grasso and Sergei M. Kuzenko}
\vspace{5mm}

\footnotesize{
{\it Department of Physics M013, The University of Western Australia\\
35 Stirling Highway, Perth W.A. 6009, Australia}}
\vspace{2mm}
~\\
Email: \texttt{darren.grasso@uwa.edu.au, sergei.kuzenko@uwa.edu.au}\\
\vspace{2mm}

\end{center}

\begin{abstract}
\baselineskip=14pt
\noindent
We study the quantum dynamics of a system of $n$ Abelian ${\cal N}=1$ vector multiplets coupled to $\frac 12 n(n+1)$ chiral multiplets which parametrise the Hermitian symmetric space
$\mathsf{Sp}(2n, {\mathbb R})/ \mathsf{U}(n)$. In the presence of supergravity, this model is super-Weyl invariant and possesses the maximal non-compact duality group $\mathsf{Sp}(2n, {\mathbb R})$ at the classical level.
These symmetries should be respected by the logarithmically divergent term (the ``induced action'') of the effective action obtained by integrating out the vector multiplets. In computing the effective action, one has to deal with non-minimal operators for which the known heat kernel techniques are not directly applicable, even in flat (super)space. In this paper we develop a method to compute the induced action in Minkowski superspace. The induced action is derived in closed form and has a simple structure. It is a   higher-derivative superconformal sigma model on $\mathsf{Sp}(2n, {\mathbb R})/ \mathsf{U}(n)$.
The obtained ${\cal N}=1$ results are generalised to the case of ${\cal N}=2$ local supersymmetry: a system of $n$ Abelian ${\cal N}=2$ vector multiplets coupled to ${\cal N}=2$ chiral multiplets $X^I$ parametrising $\mathsf{Sp}(2n, {\mathbb R})/ \mathsf{U}(n)$.
The induced action is shown to be proportional to
$ \int
{\rm d}^4x {\rm d}^4 \theta  {\rm d}^4 \bar \theta \,
E \, {\mathfrak K}(X, \bar X )$, where ${\mathfrak K}(X, \bar X )$ is the K\"ahler potential for $\mathsf{Sp}(2n, {\mathbb R})/ \mathsf{U}(n)$.
We also apply our method to compute DeWitt's $a_2 $ coefficients in some non-supersymmetric theories with non-minimal operators.
\end{abstract}
\vspace{1cm}

\vfill
\end{titlepage}

\tableofcontents{}
\vspace{1cm}
\bigskip\hrule

\allowdisplaybreaks


\section{Introduction}\label{Section1}

Four-derivative quantum corrections are ubiquitous in quantum field theory in curved space \cite{DeWitt,BirrellDavies,BV,FT85,BOS}. In (super)conformal field theories, the one-loop logarithmically divergent parts of the effective action are (super)conformal.
Recently, a family of  higher-derivative superconformal sigma models were proposed \cite{K2020} that are expected to originate as induced actions in certain supersymmetric gauge theories. Such a $\s$-model is associated
with an arbitrary K\"ahler manifold  ${\mathfrak M}^{n}$, with $n$ the complex dimension. In a background of $\cN=1$ supergravity, the $\s$-model is described by
$n$ covariantly chiral scalar superfields $\F^I$, $\bar \cD_\ad \F^I =0$, and their conjugates
$\bar \F^{\bar I}$, which parametrise  ${\mathfrak M}^{n}$. The action is given by
\bea
S &=& \frac{1}{16} \int
\rd^{4|4}z\,
E \, \bigg\{
{\mathfrak g}_{I \bar J} (\F, \bar \F)
\Big[ \nabla^2 \F^I \bar \nabla^2 \bar \F^{\bar J}
-8 G_{\a\ad}\cD^\a \F^I  \bar \cD^{\ad} \bar \F^{\bar J} \Big] \non \\
&& \qquad
+{\mathfrak F}_{IJ \bar K \bar L} (\F, \bar \F)
 \cD^\a \F^I \cD_\a \F^J \bar \cD_\ad
\bar \F^{\bar K} \bar \cD^\ad \bar \F^{\bar L} \bigg\}
~,
\label{sigma}
\eea
where $G_{\a\ad}$ is one of the superspace torsion tensors
in the Grimm-Wess-Zumino geometry \cite{GWZ}
(we follow the supergravity conventions of  \cite{Buchbinder:1998qv,KRT-M}),
${\mathfrak g}_{I\bar J} = \pa_I \pa_{\bar J} {\mathfrak K}$ is the K\"ahler metric,
\bea
\nabla^2 \F^I = \cD^2 \F^I + \G^I_{KL} \cD^\a \F^K \cD_\a \F^L~,
\label{nabla2}
\eea
and ${\mathfrak F}_{IJ \bar K \bar L} $ is  a tensor field on the
target space that is constructed from the K\"ahler metric ${\mathfrak g}_{I\bar J}$,
Riemann tensor ${\mathfrak R}_{I \bar J K \bar L} $ and, in general,
its covariant derivatives.\footnote{We recall that the Christoffel symbols $\G^I_{KL} $ and the curvature
${\mathfrak R}_{I \bar J K  \bar L} $ are given by the expressions
$
\G^I_{JK} = g^{I \bar L} \pa_J \pa_K \pa_{\bar L} {\mathfrak K}$ and
$ {\mathfrak R}_{I \bar J K  \bar L} = \pa_I \pa_K \pa_{\bar J} \pa_{\bar L}  {\mathfrak K}
-{\mathfrak g}^{M \bar N} \pa_I \pa_K \pa_{\bar N} {\mathfrak K}  \pa_{\bar J} \pa_{\bar L}  \pa_M {\mathfrak K}$.}
A typical expression for $ {\mathfrak F}_{IJ \bar K \bar L}$ is
\bea
 {\mathfrak F}_{IJ \bar K \bar L} =  \a_1 {\mathfrak R}_{(I \bar K J)\bar L}
 +\a_2 {\mathfrak g}_{(I \bar K} {\mathfrak g}_{J) \bar L} +\dots~,
 \label{F-structure}
 \eea
 with $\a_1$ and $\alpha_2$ numerical coefficients.
It may be shown that the action \eqref{sigma} is super-Weyl invariant provided $\F^I$ is  inert under the super-Weyl transformations.\footnote{The super-Weyl invariance of  \eqref{sigma} may be traced  to the existence of the $\cN=1$ superconformal operator constructed in \cite{BdeWKL}. That operator is a unique $\cN=1$ supersymmetric extension of the conformal fourth-order operator,
$\Delta_0 = (\nabla^a \nabla_a)^2 + 2 \nabla^a \big( {R}_{ab} \,\nabla^b
	- \tfrac{1}{3} {R} \,\nabla_a	\big)$,
proposed for the first time by Fradkin and Tseytlin \cite{FT1982}.}
In a Minkowski superspace background, the model \eqref{sigma} is superconformal.

It was demonstrated in \cite{K2020} that a special case of \eqref{sigma} emerges as an induced action in the model for a massless vector multiplet coupled to a dilaton-axion chiral superfield $\F$, $\bar \cD^\ad \F=0$.
Its classical dynamics is described by the action
\bea
S [V; \F , \bar \F]= - \frac{\ri }{4} \int \rd^4x \rd^2 \q  \, \cE \, \F W^\a W_\a +{\rm c.c.}~,
\qquad W_\a = -\frac 14 (\bar \cD^2 -4R) \cD_\a V~.
\label{VM-action}
\eea
The chiral field strength $W_\a$ and the action are invariant under gauge transformations
\bea
\d_\l V =\l + \bar \l ~, \qquad \bar \cD^\ad \l =0~.
\eea
The gauge prepotential $V $  and the dilaton-axion superfield $\F$ are super-Weyl inert,
which implies that the action \eqref{VM-action} is super-Weyl invariant.
In addition, the model possesses $\sSL(2,{\mathbb R})$ duality \cite{KT2}.
The duality group acts on $\F$ by fractional linear transformations
\bea
\F \to \F' = \frac{a\F + b}{c\F +d} ~, \qquad \left(
\begin{array}{cc}
 a\quad  & b\\
c\quad &   d
\end{array}
\right) \in \sSL(2,{\mathbb R}) ~.
\label{fractional_linear}
\eea
The chiral scalar superfield $\F$ and its conjugate $\bar \F$ parametrise  the Hermitian symmetric space
$\sSL(2,{\mathbb R}) / \sSO(2) $ with metric
\bea
\rd {\mathfrak s}^2 = -\frac{4 \,\rd \bar \F\, \rd  \F}{ ( \bar \F - \F)^2 } \quad \Longrightarrow \quad
\G^\F_{\F \F} = \frac{2}{\bar \F - \F} ~, \quad
{\mathfrak R}_{\F \bar \F \F \bar \F} =  \frac{8}{(\bar \F - \F)^4} ~.
\eea

One of the aims of the present paper is to extend the analysis of \cite{K2020} to a model for $n$ Abelian vector multiplets  $V_i$,  with $i=1,\ldots,n$, coupled to
$\frac 12 n(n+1)$  chiral scalar superfields $\F^{ij}  = \F^{ji}$
and their conjugates $\bar \F^{\bar i\bar j } = \bar \F^{\bar j \bar i}$
parametrising the homogeneous space $\sSp(2n,\dsR)/\sU(n)$,
\begin{equation}
  \F=(\F^{ij}) =\F^{\rT}\in \sMat (n, {\mathbb C})
	~,\qquad \ri(\bar \F -\F)>0~.
	\label{eqn:DilatonAxionConstraints}
\end{equation}
Its classical action is a natural generalisation of \eqref{VM-action}
\begin{align}
  S[V;\F, \bar{\F}]= -\frac{\ii}{4}\int\!
     \rd^4x \rd^2 \q \, \cE
  \, W_i^\a \F^{ij} W_{\a j} +\mathrm{c.c.}\,,  \label{ca}
\end{align}
where the $n$ chiral field strengths $W_{\a i}$ are defined similarly to  \eqref{VM-action}.
By construction, the model under consideration is super-Weyl invariant. In addition, it can be seen to possess the maximal possible duality group, $\sSp(2n, {\mathbb R})$, see  \cite{KT2} for the technical details. The logarithmically divergent part of the effective action, which is obtained by integrating out the vector multiplets and
is referred to as the induced action, must respect these properties.
As we will see, the main technical problem with computing the effective action is that we have to deal with non-minimal operators for which the known heat kernel techniques are not directly applicable, even in flat superspace. In order to concentrate on addressing this fundamental challenge and avoid the additional complications arising from curved superspace, we simplify our analysis by restricting the model to Minkowski superspace.

Rearranged and cast on full Minkowski superspace the above action becomes
\begin{align}
  S[V;\F, \bar{\F}]= \frac{1}{16}\int\! \dd^{4|4}z\, V\Big(\X D^\a \DB^2 D_\a
  +  (D^\a \X) \DB^2 D_\a + (\DB_\ad \X)D^2 \DB^\ad  \Big)V\,, \label{ca2}
\end{align}
where here, and henceforth, we suppress matrix indices
and make use of the positive definite symmetric matrix
\begin{align}
\Xi := \ii(\ba{\F}-\F)~.
\label{111}
\end{align}
Our goal is
to compute the logarithmically divergent part of the effective action, $\G [\F, \bar \F]$, defined by integrating out the vector multiplets
\begin{align}
{\rm e}^{\ri \G [\F, \bar \F]} &=\int [{\mathfrak D} V] \,
\d_+ \big[\k (V) \big] \, \d_{-} \big[\ba{\k} (V) \big] (\rm Det \,H)\,\re^{\ri S[V; \F, \bar \F] }\,,
\label{ee}
 \end{align}
where $\d_{\pm}$ denotes functional (anti)chiral delta functions, $\k(V)$ is a gauge fixing condition, and $H$ the Faddeev-Popov operator.  We choose the useful gauge fixing condition
\begin{align}
  \k(V) =  -\frac14 \DB^2 V + \eta \,, \qquad  \DB_\ad \eta =0\,,
\end{align}
where $\eta =(\eta_i)$ is a collection of arbitrary background chiral superfields, which gives rise to the following Faddeev-Popov operator
\begin{align}
  H = \left(
         \begin{array}{cc}
           0 & -\frac14 \DB^2 \\
           -\frac14 D^2 & 0 \\
         \end{array}
       \right)\,.
\end{align}
The operator $H$ is independent of the background fields $\F$ and $\bar{\F}$, and so will be ignored from here onward.

The right-hand side of the effective action \eqref{ee} is independent of the superfields $\eta$ and $\ba{\eta}$ and so we can may integrate over them with the following weight:
\begin{align}
  ({\rm Det}H_\X)^\frac12\ex^{-\ii S_{\rm GF}[\eta,\ba{\eta};\X]}\,,
\end{align}
where $S_{\rm GF}$ ultimately becomes our gauge fixing term, which we choose to be
\begin{align}
S_{\textrm{GF}}[\eta,\ba{\eta};\X] = \int\! \dd^{4|4}z\, \ba{\eta}\, \X \,\eta\,,
\end{align}
and $({\rm Det}H_\X)^\frac12$ is the Neilson-Kallosh ghost operator
\begin{align}
  H_\X = \left(
         \begin{array}{cc}
           0 & -\frac14 \DB^2 \X \\
           -\frac14 D^2 \X & 0 \\
         \end{array}
       \right)
\end{align}
defined to act on the space of column vectors
\begin{align}
       \left(
         \begin{array}{c}
            \eta\\
          \ba{\eta} \\
         \end{array}
       \right)\,, \qquad  \DB_\ad \eta =0\,,
\end{align}
such that
\begin{align}
     H_\X  \left(
         \begin{array}{c}
            \eta\\
          \ba{\eta} \\
         \end{array}
       \right) =  \left(
         \begin{array}{c}
           -\frac14 \DB^2 (\X \ba{\eta})\\
          -\frac14 D^2 (\X \eta)   \\
         \end{array}
       \right) \,.
\end{align}
The above quantisation procedure leads to the following representation of the effective action
\begin{align}
{\rm e}^{\ri \G [\F, \bar \F]} &=({\rm Det}H_\X)^\frac12 \int [{\mathfrak D} V] \,
\ex^{\ii S_\mathrm{tot}}\,,
\label{ee2}
\end{align}
where the total action,  $S_\mathrm{tot}$, is the sum of the classical action and the gauge fixing term:
\begin{align}
  S_\mathrm{tot} &=  S[V;\F, \bar{\F}]-  \frac{1}{16} \int\! \dd^{4|4}z\,  (\DB^2 V) \X  (D^2 V) = -\frac12 \int\! \dd^{4|4}z\, V \op_\mathrm{v} V\,.
\end{align}
Here the vector operator $\op_\mathrm{v}$ is defined by
\begin{align}
  \op_\mathrm{v} = \X \Box  +\frac{1}{16}(D^2 \X) \DB^2 +\frac{1}{16}(\DB^2 \X) D^2 -\frac{\ii}{2} (D^\a \X) \pd_{\a\ad}\DB^{\ad}-\frac{\ii}{2} (\DB^\ad \X) \pd_{\a\ad}D^{\a}\,, \label{vectoroperator}
\end{align}
and $\Box=\pd^a\pd_a$ is the d'Alembertian.

The effective action \eqref{ee2} therefore becomes
\begin{align}
  \G [\F, \bar \F] = \frac{\ii}{2} \Tr \ln \op_\mathrm{v}  - \frac{\ii}{2} \Tr \ln  H_\X\,.
\end{align}
To facilitate computing $\Tr \ln  H_\X$ we employ the so-called `doubling trick'
 \begin{align}
   \Tr \ln  H_\X  =  \frac12 \Tr \ln  (H_\X^2)\,,
 \end{align}
where
\begin{align}
     H_\X^2  = \left(
         \begin{array}{cc}
           \frac{1}{16} \DB^2 \X D^2 \X & 0 \\
           0 & \frac{1}{16} D^2 \X \DB^2 \X \\
         \end{array}
       \right)= \left(
         \begin{array}{cc}
           \op_\mathrm{+} & 0 \\
           0 & \op_\mathrm{-} \\
         \end{array}
       \right)
\end{align}
and so
\begin{align}
  \G [\F, \bar \F] = \frac{\ii}{2} \Tr \ln \op_\mathrm{v}  - \frac{\ii}{4} \Tr_+ \ln  \op_\mathrm{+} - \frac{\ii}{4} \Tr_{-} \ln  \op_\mathrm{-}\,. \label{ee3}
\end{align}
In the expressions above we have defined (anti)chiral operators $\op_\mathrm{\pm}$
\begin{align}
  \op_{+}:=  \frac{1}{16} \DB^2 \X D^2 \X\,,  \qquad\qquad  \op_\mathrm{-}:=  \frac{1}{16} D^2 \X \DB^2 \X\,,
\end{align}
and  $\Tr_+$ denotes the chiral functional trace
\begin{align}
  \Tr_{+} P = \int \!
 \rd^4x \rd^2 \q
   \,\tr P(z,z)\,, \qquad \qquad P(z,z'):= P \mathbbm{1}_n \d_+(z,z')\,,
\end{align}
of an operator $P$ acting on the space of chiral scalar superfields, where $\mathbbm{1}_n$ is the $n\times n$ identity matrix, `tr' denotes the trace over matrix indices, and
$\d_+(z,z')$ the chiral delta-function
\begin{align}
  \d_+(z,z') = -\frac{1}{4}\DB^2 \d^{(4|4)}(z,z')\,, \qquad  \d^{(4|4)}(z,z') =  \d^{(4)}(x,x') \d^{(2)}(\q- \q') \d^{(2)}(\ba{\q}-\ba{\q}')\,.
\end{align}
The action of $\op_\mathrm{+}$ on chiral scalar fields $\eta$ is found to be
\begin{align}
  \op_\mathrm{+}\eta &=  \Big(\X^2 \Box -\frac{\ii}{2}(\DB^{\ad}\X^2)\pd_{\a\ad}D^{\a} +\frac{1}{16} (\DB^2\X^2) D^2 \qquad\qquad \no \\ &\qquad\qquad -\frac{\ii}{2}\DB^{\ad}(\X D^\a \X)\pd_{\a\ad}+\frac18 \DB^2(\X D^\a \X)D_\a +\frac{1}{16} \DB^2(\X D^2 \X)\Big) \eta\,.
\end{align}

Computing the logarithmically divergent part of the effective action \eqref{ee3} using heat kernel techniques now amounts to determining the trace of the diagonal $a_2$ DeWitt coefficients associated with the operators $\op_\mathrm{v}$  and $\op_\mathrm{\pm}$.  As mentioned, this is not a straight forward exercise since the standard superfield
Schwinger-DeWitt techniques \cite{Buchbinder:1998qv}
are not applicable here, due to the fact that  $\op_\mathrm{v}$  and $\op_\mathrm{\pm}$ are non-minimal second-order operators of the general form $\mathcal{O}= M \Box + \cdots$, where the coefficient $M$ of the d'Alembertian is a matrix valued field.  Although the literature on minimal operators is extensive, there are  fewer publications on non-minimal operators, dating back only to the early 1990's
\cite{Gusynin2,Gusynin3,Gilkey} (for a comprehensive list of references see \cite{Barvinsky:2021ijq}).  More recently there have been a number of works devising techniques which can be used to compute DeWitt coefficients for non-minimal operators -- for example see \cite{Barvinsky:2021ijq,Iochum:2016ynh,Iochum:2017ver,Leung:2019bgo} and references therein -- however, as far as we know, no explicit calculations have been performed which generate results for our case of interest here, namely the $a_2$ coefficient for non-minimal operators where $M$ is a general positive definite matrix valued function.  The approach and analysis of \cite{Iochum:2016ynh,Iochum:2017ver} appears generally applicable, however it requires an intermediate spectral decomposition of the matrix coefficient $M$, which at the end of the computation needs to be reassembled back into powers of the full matrix and its inverse, a non-trivial task.  In \cite{Leung:2019bgo} a method similar to that of \cite{Iochum:2016ynh,Iochum:2017ver} is used in curved $\cN=1$ superspace to compute the one-loop divergence of the dilaton-coupled super Yang-Mills theory, however the coefficient $M$ in that case is not matrix valued.  Additionally, although the work \cite{Barvinsky:2021ijq} does address non-minimal operators and also appears generally applicable, their method does not provide a closed form expression in terms of a  general matrix coefficient $M$.  In particular, if used directly on the non-minimal operators of interest here the approach of \cite{Barvinsky:2021ijq} yields infinite summations involving commutators of $M$.

In the present situation, one may compute the $a_2$ coefficient associated with the operator $\op_\mathrm{v}$ by invoking a field redefinition in the path integral -- for example one may work with the operator $\X^{-\frac12}\op_\mathrm{v}\X^{-\frac12}$ -- leaving a minimal operator for which standard techniques apply.\footnote{For example, this was the procedure used in the recent work \cite{GKP} which computed the non-supersymmetric version of the theory of interest here.}    However, no such field redefinition is possible in the case of the (anti)chiral operators $\op_\mathrm{\pm}$ and so another approach must be used which is applicable to the more general case of non-minimal operators.  Following the work of \cite{Iochum:2016ynh,Iochum:2017ver}, here we devise a technique that allows us to directly compute the trace of the diagonal $a_2$ coefficient (up to integration by parts) for the class of non-minimal operators in which we are interested, without the need for an intermediate spectral decomposition.

This paper is structured as follows.  First, as a means of introducing our approach for computing the first few DeWitt coefficients of second-order non-minimal operators, in section \ref{sec:Minkowskispace} we explain the technique in the conceptually simpler situation of Minkowski space.  Then, as an immediate example of this approach, in section \ref{sec:app} we apply it to the non-supersymmetric version of the model \eqref{ca2}, finding agreement with the results of the recent work \cite{GKP} in the absence of gravity.  In sections \ref{sec:superspace} and \ref{sec:chiral} we then respectively extend the approach to operators of the form $\op_\mathrm{v}$  and $\op_\mathrm{\pm}$ in Minkowski superspace.  In section \ref{sec:final} we conclude with our final ${\cal N}=1$ results and explain how they may be generalised to the case of ${\cal N}=2$ local supersymmetry.  Three appendices appear in this paper. Appendices \ref{appA} and \ref{proof3} contain the details of some proofs, and appendix \ref{appB} contains the full expression for the $a_2$ coefficient associated with a general chiral operator of the form $\op_\mathrm{+}$.

\section{Heat kernel coefficients of non-minimal operators in Minkowski space}\label{sec:Minkowskispace}

Here we consider non-minimal operators in four-dimensional flat spacetime of the general form
\begin{align}
\op = M\Box +V^a \pd _a +T\,, \label{op}
\end{align}
with $M=M(x)$, $V^a=V^a(x)$ and $T=T(x)$ all $n \times n$ matrix valued and $M=M^\textrm{T}$ positive definite.\footnote{Positive definiteness is not actually required, but it suffices for our purposes here.} We note that second-order operators of this kind typically emerge from underlying theories involving gauge covariant derivatives (rather than partial derivatives), however such operators can be brought into the above form, which turns out to be best adapted to our approach.

The heat kernel $K(x,x';s)$ associated with \eqref{op} is defined as follows
\begin{align}
K(x,x';s) = \ex^{\ii s \op }\mathbbm{1}_n\d^{(4)}(x-x')\,.
\end{align}
Using an integral representation  of the delta function\footnote{Strictly speaking the delta-function should be accompanied by a factor $\mathcal{I}(x,x')$ such that $\mathcal{I}(x,x)=\mathbbm{1}_n$, included to ensure the correct gauge transformation properties of the heat kernel.  In our case of interest -- diagonal heat kernel coefficients -- this factor may be ignored.}
\begin{equation}\label{eq:delta}
\d^{(4)}(x-x')=\int\!  \dd k \, \re^{\ii
k_a(x^a-x'^a)}\,, \qquad \qquad \dd k:  = \frac{\dd^4 k}{(2 \p)^4}\,,
\end{equation}
the heat kernel becomes
\begin{align}
K(x,x';s) = \int\!  \dd k \,  \ex^{\ii s \op } \re^{\ii k_a(x^a-x'^a)}  = \int\!  \dd k \,  \re^{\ii k_a(x^a-x'^a)} \ex^{\ii s \hat{\op} }\,,
\end{align}
where
\begin{align}
\hat{\op}  = M X^a X_a +V^a X_a +T\,, \qquad\qquad X_a=\pd_a + \ii k_a\,.
\end{align}
For the purposes of computing contributions to an effective action, we are interested in the functional trace of the heat kernel which is given by
\begin{align}
K(s) = \int\! \dd^4 x  K(x;s) \label{k0}\, ,
\end{align}
where
\begin{align}
K(x;s) = \lim_{x' \rightarrow x} \tr\, K(x,x';s) = \tr \int\! \dd k\, \ex^{\ii s \hat{\op} }\,.  \label{kk}
\end{align}
The kernel $K(x;s)$ has the well-known asymptotic expansion
\begin{align}
K(x;s) = \frac{h}{s^2}\sum_{n=0}^{\infty}(\ii s)^n\, a_n(x)\,,
\end{align}
where $h= \ii /(4 \p \ii )^2$, and $a_n(x)$ denotes the matrix trace of the DeWitt heat kernel coefficients in the coincidence limit.

For the purpose of identifying the various heat kernel coefficients it is useful to make the $k$ dependance of the operator $\hat{\op}$ explicit, and write
\begin{align}
\hat{\op}  = \op +k^a G_a -k^2 M \,, \qquad \qquad k^2 = k^a k_a\,,
\end{align}
where
\begin{align}
  G_a= 2 \ii M \pd_a + \ii V_a\,.
\end{align}
After the rescaling $k_a \rightarrow k_a/\sqrt{s}$, expression \eqref{kk} becomes
\begin{align}
K(x;s) = \frac{\tr}{s^2}\int\! \dd k \, \ex^{-\ii k^2 M + \ii s \op + \ii \sqrt{s}k^a G_a}\,. \label{k2}
\end{align}
We now expand the above expression using the Dyson series-like expansion\footnote{Formally known as the Volterra series, for example see \cite{Avramidi}.}
\begin{align}
\ex^{A+B}= \ex^A + \ex^A \sum_{n=1}^{\infty} \int_{0}^{1} \! \dd y_1\int_{0}^{y_1} \! \dd y_2 \ldots \int_{0}^{y_{n-1}}\! \dd y_{n} \left(\ex^{-y_1 A}B\ex^{y_1 A}\right)&\left(\ex^{-y_2 A}B\ex^{y_2 A}\right) \times \ldots  \no \\ \ldots  &\times \left(\ex^{-y_n A}B\ex^{y_n A}\right)\,,
\end{align}
where
\begin{align}
  A = -\ii k^2 M\,, \qquad \qquad B =\ii s \op + \ii \sqrt{s}k^a G_a\,,
\end{align}
with $\ex^A$ providing the convergence factor.

Adopting a notation similar to that introduced in \cite{Iochum:2016ynh} (also used in \cite{Iochum:2017ver} and \cite{Leung:2019bgo}), we define
\begin{align}
H_n [B_1 \otimes B_2 \otimes \cdots \otimes B_n]:= \ex^A \int_{0}^{1} & \! \dd y_1\int_{0}^{y_1} \! \dd y_2 \ldots \int_{0}^{y_{n-1}}\! \dd y_{n} \left(\ex^{-y_1 A}B_1\ex^{y_1 A}\right) \no \\ &\times \left(\ex^{-y_2 A}B_2\ex^{y_2 A}\right) \times \ldots \times \left(\ex^{-y_n A}B_n\ex^{y_n A}\right)\,, \label{defHn}
\end{align}
and so the heat kernel $K(x;s)$ becomes
\begin{align}
K(x;s) = \frac{\tr}{s^2}\int\! \dd k\,\bigg\{\ex^A + \sum_{n=1}^{\infty} H_n [B \otimes B \otimes \cdots \otimes B]\bigg\}\,. \label{hke}
\end{align}
To compute the DeWitt coefficients up to $a_2$ we must include terms up to order $s^0$ in \eqref{hke}:
\begin{align}
K(x;s) &= \frac{\tr}{s^2}\int\! \dd k \,\ex^{-\ii k^2 M}
+\frac{\tr}{s}\int\! \dd k \bigg\{ \ii H_1 [\op] - k_a k_b  H_2 [G^a \otimes G^b]\bigg\}  \no  \\
&+\tr\int\! \dd k \bigg\{ - H_2 [\op \otimes \op] -  \ii k_a k_b \bigg(H_3 [\op \otimes G^a \otimes G^b]+H_3 [ G^a \otimes \op \otimes G^b] \no \\&+H_3 [ G^a \otimes G^b \otimes \op ]\bigg) + k_a k_b k_c k_d  H_4 [G^a \otimes G^b \otimes G^c \otimes G^d]\bigg\} + \mathcal{O}(s^1)\,.
\end{align}
In this expression, and from here onward, all of the arguments of the functions $H_n$ are understood as being independent of $k$.  As anticipated, no fractional powers of $s$ appear in the above series due to the fact that such terms are odd in $k$ and therefore vanish under the integral.  We now use the above expression to compute the first three DeWitt coefficients in turn.

\subsection{The $a_0$ contribution}

The $a_0$ contribution to the heat kernel, denoted $K(x;s)\big|_{a_0}$,  is simply
\begin{align}
K(x;s)\Big|_{a_0} &= \frac{\tr}{s^2}\int\! \dd k\, \ex^{-\ii k^2 M} = \frac{h}{s^2} \tr \,M^{-2}
\quad \implies \quad a_0 =  \tr\, M^{-2}\,,
\end{align}
having used the identity
\begin{align}
\int\! \dd k\, (k^2)^n \ex^{-\ii k^2 M} = (-\ii)^n (n+1)! \,h\,  M^{-(n+2)} \qquad \qquad n\geq 0\,,  \label{knid}
\end{align}
which may be established by diagonalising $M$.

\subsection{The $a_1$ contribution}

The $a_1$ contribution of the heat kernel, $K(x;s)\big|_{a_1}$,  is
\begin{align}
K(x;s)\Big|_{a_1} &= \frac{\tr}{s}\int\! \dd k \bigg\{ \ii H_1 [\op] - k_a k_b  H_2 [G^a \otimes G^b]\bigg\} \,.
\end{align}
Noting that each $H_n [B_1 \otimes B_2 \otimes \cdots \otimes B_n]$ contains the product of the convergence factor $\exp(-\ii k^2 M)$ and a power series of positive powers of $k^2$, we use
\begin{align}
\int\! \dd k \, k_a k_b (k^2)^n \ex^{-\ii k^2 M}  =\frac{\h_{ab}}{4} \int\! \dd k\, (k^2)^{n+1}\ex^{-\ii k^2 M}  \qquad \qquad n\geq 0 \,, \label{kprod}
\end{align}
with $\h_{ab}$ the Minkowski metric, to write
\begin{align}
K(x;s)\Big|_{a_1} &= \frac{\tr}{s}\int\! \dd k \bigg\{ \ii H_1 [\op] - \frac{k^2}{4} H_2 [G^a \otimes G_a]\bigg\} \,. \label{a1}
\end{align}
Expressions like \eqref{kprod} themselves may be established inductively by using \eqref{knid} and what we will refer to as a \emph{Gaussian moment generating identity}:
\begin{align}
0=\int\! \dd k \, \frac{\pd}{\pd k_a}\Big(k_b (k^2)^n \,\ex^{-\ii k^2 M}\Big)\,. \label{gid}
\end{align}

The coefficient $a_1$ turns out to be unique amongst the higher order DeWitt coefficients in that it may be directly computed with very little effort as follows.

First we define the operator $L_A(B)=[A,B]$, or more explicitly
\begin{align}
  L_A^0(B)=B\,, \qquad \qquad L_A^n(B)=[A,L_A^{n-1}(B)]\,,
\end{align}
and then using the following identity for inverse powers of the matrix $M$,
\begin{align}
M^{-m} =\frac{1}{(m-1)!}\int_{0}^{\infty}\dd z\, z^{m-1}\ex^{-zM}  \qquad \qquad  m \geq 1\,,
\end{align}
we establish that, for any operator $P$,
\begin{align}
P M^{-m} &= \frac{1}{(m-1)!}\int_{0}^{\infty}\dd z\, z^{m-1}P \ex^{-zM} \no \\
&= \frac{1}{(m-1)!}\int_{0}^{\infty}\dd z\, z^{m-1}\ex^{-zM}\left(\ex^{zM}P \ex^{-zM}\right) \no \\
&= \frac{1}{(m-1)!}\int_{0}^{\infty}\dd z\, z^{m-1}\ex^{-zM}\sum_{n=0}^{\infty}\frac{z^n}{n!}\,L_M^n (P)\,,
\end{align}
and so
\begin{align}
P M^{-m}= \sum_{n=0}^{\infty}\frac{(n+m-1)!}{n!(m-1)!} M^{-(n+m)} L_M^n (P)\,.  \label{PMid}
\end{align}

The first term in \eqref{a1} then becomes
\begin{align}
 \int\! \dd k\,  H_1 [\op]   &=\int\! \dd k\, \ex^A \int_{0}^{1}\! dy\,\ex^{-y A}\op\ex^{y A} \no \\
  &=\int\! \dd k\, \ex^A \sum_{n=0}^{\infty}\frac{(-1)^n}{(n+1)!}L_A^n (\op) \no \\
  &=\sum_{n=0}^{\infty}\frac{\ii^n}{(n+1)!}\int\! \dd k\, (k^2)^n \ex^{-\ii k^2 M} L_M^n (\op) \no \\
   &=h\sum_{n=0}^{\infty} M^{-(n+2)} L_M^n (\op) \no \\
    &=h M^{-1}\op M^{-1}\,, \label{a0msa}
\end{align}
having used \eqref{knid} and \eqref{PMid}\,.

Similarly, the second term in \eqref{a1} becomes
\begin{align}
\int\! \dd k\,  k^2 H_2 [G^a \otimes G_a] &=\int\! \dd k\, k^2 \ex^A \int_{0}^{1} \! dy\int_{0}^{y} \! dz \,(\ex^{-y A}G^a\ex^{y A})(\ex^{-z A}G_a\ex^{z A}) \no \\
&=\int\! \dd k\, k^2 \ex^A \sum_{m=0}^{\infty}\sum_{n=0}^{\infty} \frac{(-1)^{m+n}}{n!(m+1)!(n+m+2)}L_A^n (G^a)L_A^m (G_a)\no\\
&=- \ii h \sum_{m=0}^{\infty}\sum_{n=0}^{\infty} \frac{(n+m+1)!}{n!(m+1)!}M^{-(n+m+3)}L_M^n (G^a)L_M^m (G_a)\no \\
&=- \ii h M^{-1}G^a M^{-1}G_a M^{-1}\,, \label{a0msb}
\end{align}
where in the last line we have used the following result (established by using \eqref{PMid} twice)
\begin{align}
P M^{-1}Q M^{-1}= \sum_{m=0}^{\infty}\sum_{n=0}^{\infty} \frac{(n+m+1)!}{n!(m+1)!}M^{-(n+m+2)}L_M^n (P)L_M^m (Q)
\end{align}
for arbitrary operators $P$ and $Q$.

It follows that the entire $a_1$ contribution to the heat kernel  is
\begin{align}
K(x;s)\Big|_{a_1} &= \frac{\tr}{s}\bigg\{ \ii h M^{-1}\op M^{-1}+ \frac{\ii h }{4}  M^{-1}G^a M^{-1}G_a M^{-1} \bigg\} \,.
\end{align}
Expanding and acting through to the right with the differential operators we immediately generate the following result:
\begin{align}
a_1 = \tr\bigg( M^{-1}T M^{-1}+ \frac12 M^{-1}(\pd^a M)M^{-1}V_a M^{-1} \qquad\qquad\qquad\qquad\qquad\qquad \no \\ \qquad\qquad\qquad- \frac12 M^{-1}(\pd^a V_a) M^{-1}-\frac14 M^{-1}V^a M^{-1}V_a M^{-1}\bigg)\,.
\end{align}

We note that the above result agrees with that provided in  \cite{Iochum:2016ynh} and \cite{Iochum:2017ver}, but the approach presented here differs from theirs.  Although far more general in scope, their approach involves a spectral decomposition of $M$ to compute the terms in the Dyson series, and then at the end of the calculation one reassembles the result into factors of $M^{-1}$.  Computing $a_2$ is significantly more challenging than computing $a_1$, but we again find that we can avoid decomposing $M$ at the cost of integrating by parts.

\subsection{The $a_2$ contribution}

The $a_2$ contribution to the heat kernel, $K(x;s)\big|_{a_2}$,  is
\begin{align}
K(x;s)\Big|_{a_2} &=\tr\int\! \dd k \bigg\{ - H_2 [\op \otimes \op] -  \frac{\ii k^2}{4} \bigg(H_3 [\op \otimes G^a \otimes G_a]+H_3 [ G^a \otimes \op \otimes G_a] \no \\ &\qquad+H_3 [ G^a \otimes G_a \otimes \op ]\bigg) + \frac{(k^2)^2}{4!} R_{abcd} H_4 [G^a \otimes G^b \otimes G^c \otimes G^d]\bigg\}\,, \label{a2}
\end{align}
where we have used \eqref{kprod} and the following identity
\begin{align}
\int\! \dd k \, k_a k_b k_c k_d (k^2)^n \ex^{-\ii k^2 M} =\frac{R_{abcd}}{4!} \int\! \dd k  \, (k^2)^{n+2} \ex^{-\ii k^2 M} \qquad  \qquad n\geq 0\,, \label{kprod2}
\end{align}
where $R_{abcd}$ is the totally symmetric tensor
\begin{align}
  R_{abcd}= \h_{ab}\h_{cd}+\h_{ad}\h_{bc}+\h_{ac}\h_{bd}\,.
\end{align}
Expression \eqref{kprod2} may be established by using a Gaussian moment generating identity in a manner similar to that used to establish \eqref{kprod}.

Experience shows that to compute $a_2$ we cannot proceed directly as we did for $a_1$, since the integrals which appear in $a_2$ are all of the form
\begin{align}
\int\! \dd k \, (k^2)^{(n-2)} H_n [B_1 \otimes \cdots \otimes B_n] \qquad \qquad n \geq 2\,.  \label{int}
\end{align}
As demonstrated in \cite{Iochum:2016ynh,Iochum:2017ver} such integrals cannot, by themselves, be written explicitly in closed form terms of powers of $M$.  In fact, in those publications it was shown that integrals of this kind contain the logarithm of eigenvalues of $M$.

Instead of attacking integrals like \eqref{int} in isolation, we follow a different path where the ultimate goal is to use a collection of manipulations to recast all terms that appear in $a_2$ into \emph{cyclic combinations} of the following form:\footnote{The precise definition of this expression will be given later.}
\begin{align}
\int\! \dd k \, (k^2)^{(n-2)}\bigg\{ H_n [B_1 \otimes \cdots \otimes B_n]+ \textrm{cyclic perms of $B_i$}\bigg\}\,,
\end{align}
where the $B_i$ are $k$-independent matrices only (i.e. not differential operators).    It turns out that these cyclic combinations  can be evaluated directly using a single simple identity.  The manipulations required to generate such cyclic combinations include (clarifying examples will follow):
\begin{enumerate}[label=(\roman*)]
  \item Operating through with the derivatives originally appearing in the arguments of $H_n$ either to the right or left (so that the arguments of each $H_n$ no longer contain any differential operators, only matrices and their derivatives).  Operating with derivatives to the left is accomplished by integration by parts, permissible since we are interested in the functional trace of the heat kernel.\footnote{In this work we ignore all boundary terms, such terms do not contribute to the effective action.}
  \item Using Gaussian moment-like generating identities, analogues of expression \eqref{gid}.
\end{enumerate}

The rules governing the manipulations (i) and (ii) are established by using the following identity (Duhamel's formula) for any derivation $\pd$ and operator $P$:
\begin{align}
  (\pd \,\ex^{s P}) = \ex^{s P}  \int_{0}^{s} \! dy \, \ex^{-y P} (\pd P) \ex^{y P}\,.
\end{align}
From this it follows\footnote{The essential steps for the proofs are given in \cite{Iochum:2016ynh}.} that for matrices $B_1$, $B_2$, \ldots, $B_k$,  derivatives acting through to the right satisfy
\begin{align}
 H_k [B_1 \otimes B_2 \otimes \cdots & \otimes B_i  \pd \otimes B_{i+1}\otimes \cdots  \otimes B_k ]  = \sum_{j=i+1}^{k} H_k [B_1 \otimes B_2 \otimes \cdots \otimes (\pd B_j)  \otimes \cdots  \otimes B_k ]\no \\ &+\sum_{j=i}^{k} H_{k+1} [B_1 \otimes B_2 \otimes \cdots \otimes B_j \otimes (\pd A) \otimes B_{j+1} \otimes \cdots  \otimes B_k ]\,, \label{d1}
\end{align}
and to accommodate integration by parts -- effectively allowing us to move derivatives to the left -- we use
\begin{align}
 \pd\big(H_k [B_1 \otimes B_2 &\otimes \cdots \otimes B_k ]\big)  = H_{k+1} [(\pd A)\otimes  B_1 \otimes B_2 \otimes \cdots \otimes B_k ]
 \no \\ &+H_{k} [ (\pd B_1) \otimes B_2 \otimes \cdots \otimes B_k ]+H_{k} [ B_1\pd \otimes B_2 \otimes \cdots \otimes B_k ]\,. \label{d2}
 \end{align}

To demonstrate the details of manipulations (ii), it is sufficient to provide a single illustrative example.  Consider the identity
\begin{align}
  0 = \int\! \dd k\,  \frac{\pd}{\pd k_a}\Big(k_b H_2 [P \otimes Q ]\Big) \label{t1ex}
\end{align}
for arbitrary operators $P$ and $Q$ independent of $k$.  Since
\begin{align}
  \frac{\pd A}{\pd k_a}= -2 \ii k^a M\,,
\end{align}
then using \eqref{d1} and \eqref{d2}, and after contracting $a$ and $b$, expression \eqref{t1ex} becomes
\begin{align}
  \int\! \dd k\,  H_2 [P \otimes Q ] = \frac{\ii}{2} \int\! \dd k\, k^2 \bigg\{ H_3 [M \otimes P \otimes Q ]+H_3 [P \otimes M \otimes Q ]+H_3 [P \otimes Q \otimes M]\bigg\}\,. \label{H2toH3}
\end{align}
In general this kind of trick may  be used to relate terms of the form $(k^2)^{n-2} H_n$ to terms of the form $(k^2)^{n-1} H_{n+1}$.

As mentioned above, using the manipulations (i) and (ii), our aim is to arrange all of the terms in  $K(x;s)\big|_{a_2}$,  expression \eqref{a2}, into terms which contain only summations of $H_n$ and cyclic permutations of its $n$ matrix arguments.  To this end, for matrices $B_i$, we define:
\begin{align}
  C_n [B_1 \otimes B_2 \otimes \cdots  \otimes B_n] :=  H_n [B_1 \otimes B_2 \otimes \cdots \otimes B_n] + \textrm{cyclic perms of $B_i$}\,, \label{defcn}
\end{align}
or more precisely
\begin{align}
  C_n& [B_1 \otimes B_2 \otimes \cdots  \otimes B_n] :=  H_n [B_1 \otimes B_2 \otimes \cdots \otimes B_n] + H_n [B_n \otimes B_1 \otimes B_2 \otimes \cdots  \otimes B_{n-1}] \no \\ &+ H_n [B_{n-1} \otimes B_n \otimes B_1 \otimes \cdots  \otimes B_{n-2}] +  \cdots +  H_n [B_2 \otimes B_3 \otimes \cdots \otimes B_n \otimes B_1]\,.
\end{align}
It turns out that we can directly compute integrals of the from
\begin{align}
    I_n [B_1 \otimes B_2 \otimes \cdots  \otimes B_n] :=  \tr  \int\! \dd k\, (k^2)^{n-2} C_n [B_1 \otimes B_2 \otimes \cdots  \otimes B_n]\,,
\end{align}
where the $B_i$ are arbitrary $k$-independent matrices, using the identity
\begin{align}
    I_n [B_1 \otimes B_2 \otimes \cdots  \otimes B_n] = -(-\ii)^n h\, \tr \big( M^{-1} B_1 M^{-1} B_2 M^{-1} B_3 \cdots M^{-1} B_n \big)\,, \label{mr}
\end{align}
 or
\begin{align}
    I_n [B_1 \otimes B_2 \otimes \cdots  \otimes B_n] = -(-\ii)^n h\, \tr \big( \~B_1 \~B_2 \~B_3 \cdots  \~B_n\big)\,,
\end{align}
where in this work we define $\~P :=M^{-1}P$ for any operator $P$\,.

Identity \eqref{mr} is most directly derived as follows.  First we note that for $n\geq1$
\begin{align}
\int\! \dd k\, (k^2)^{n-1} H_n [B_1 \otimes B_2 \otimes \cdots  \otimes B_n] = (-\ii)^{n-1} h\,  M^{-1} B_1 M^{-1} B_2 M^{-1} B_3 \cdots M^{-1} B_n M^{-1}\,, \label{id}
\end{align}
which may be demonstrated directly using \eqref{PMid}  and \eqref{knid} -- see appendix A section \ref{proof1} for the details of a proof.  In viewing \eqref{mr} as a function of the matrix $M$ (treating the $B_i$ as independent of $M$), we formally differentiate with respect to $M$, $\pd/\pd M$.  In doing so we use the following rule:\footnote{More precisely, for a matrix valued function $F$ we define its derivative via $\d F = \d M_{ij}\frac{\pd F}{\pd M_{ij}}$, with $M_{ij}$ the matrix entries of $M$.  From this it follows that $\frac{\pd M_{kl}}{\pd M_{ij}} = \d_{k(i}\d_{j)l}$ where, recalling that $M$ is a symmetric matrix, the indices are symmetrised  using the bracket notation (which include a factor of 1/n!).   We then make the following identification $\frac{\pd }{\pd M} :=  \frac{\pd }{\pd M_{ii}}$ from which the rule \eqref{md} follows.}
\begin{align}
  \frac{\pd M^m}{\pd M} = m M^{m-1}  \qquad\qquad m \neq 0\,, \label{md}
\end{align}
where $M^0 := \mathbbm{1}_n$.

Using
\begin{align}
  \frac{\pd }{\pd M} \big(\ex^{-y A}B\ex^{y A}\big) =  0
\end{align}
for $B$ independent of $M$, we find:
\begin{align}
    \frac{\pd}{\pd M}\Big(I_n [B_1 &\otimes B_2 \otimes \cdots  \otimes B_n]\Big) = -\ii\, \tr  \int\! \dd k\, (k^2)^{n-1}C_n [B_1 \otimes B_2 \otimes \cdots  \otimes B_n]\no \\
    &= -\ii \, \tr  \int\! \dd k\, (k^2)^{n-1} \bigg\{H_n [B_1 \otimes B_2 \otimes \cdots \otimes B_n] + \textrm{cyclic perms of $B_i$} \bigg\} \no \\
    &= (-\ii)^n  h \,\tr  \Big(M^{-1} B_1 M^{-1} B_2 M^{-1} B_3 \cdots M^{-1} B_n M^{-1} + \textrm{cyclic perms of $B_i$}\Big) \no \\
     &= (-\ii)^n  h\, \tr  \Big(M^{-2} B_1 M^{-1} B_2 M^{-1} B_3 \cdots M^{-1} B_n \no \\
     &\qquad\qquad\qquad+ M^{-1} B_1 M^{-2} B_2 M^{-1} B_3 \cdots M^{-1} B_n  \no \\
     &\qquad\qquad\qquad+M^{-1} B_1 M^{-1} B_2 M^{-2} B_3 \cdots M^{-1} B_n + \cdots \no \\
     &\qquad\qquad\qquad+M^{-1} B_1 M^{-1} B_2 M^{-1} B_3 \cdots M^{-2} B_n\Big) \no \\
&=-(-\ii)^n  h\,\frac{\pd}{\pd M}\bigg\{ \tr\big(M^{-1} B_1 M^{-1} B_2 M^{-1} B_3 \cdots M^{-1} B_n\big) \bigg\}\,.
\end{align}
From this the result \eqref{mr} follows, at least up to terms independent of $M$, which can readily be shown to vanish by setting  $M=\mathbbm{1}_n$.  For an alternative derivation of this result using the machinery of \cite{Iochum:2016ynh} see appendix A section \ref{proof2}.

For clarity, we now provide two examples of this method.

\noindent\textbf{Example 1:} Returning to \eqref{a2}, we will compute the terms in $a_2$ which are quadratic in $T$, which we denote by $K(x;s)\big|_{a_2,T^2}$:
\begin{align}
K(x;s)\Big|_{a_2,T^2} &=\tr\int\! \dd k \Big\{ - H_2 [T \otimes T]\Big\}  =-\frac12 \tr\int\! \dd k\, C_2 [T \otimes T]  \no \\
&=-\frac12h\, \tr \big(M^{-1} T M^{-1}T\big)=-\frac12h\, \tr\, \~T^2\,.
\end{align}

\noindent\textbf{Example 2:} Now consider terms in \eqref{a2} which are linear in $T$ and $V_a$, denoted by $K(x;s)\big|_{a_2,T,V}$:
\begin{align}
K(x;s)&\Big|_{a_2,T,V} =\tr\int\! \dd k \bigg\{ - H_2 [T \otimes V^a \pd_a ] - H_2 [V^a \pd_a \otimes T] \qquad\qquad \qquad \no \\
&\quad+  \frac{\ii k^2}{2} \bigg(
H_3 [T \otimes V^a \otimes M\pd_a]
+H_3 [T \otimes M\pd_a \otimes V^a] \no \\
&\quad\qquad \qquad+H_3 [V^a \otimes T \otimes M\pd_a]
+H_3 [V^a \otimes M\pd_a \otimes T] \no \\
&\quad \qquad\qquad \qquad+H_3 [M\pd_a \otimes V^a \otimes T]
+H_3 [M\pd_a \otimes T \otimes V^a]\bigg)\bigg\}\,.
\end{align}
To generate cyclic combinations we now use \eqref{H2toH3} to push $H_2 \rightarrow k^2 H_3$, which gives the following result, where we have arranged the terms in a specific order, the reason for which will be clarified below:
\begin{align}
K(s)\Big|_{a_2,T,V} &= \frac{\ii}{2}\tr\int\! \dd k\, k^2 \bigg\{ - H_3 [M \otimes T \otimes V^a \pd_a ]- H_3 [T \otimes M \otimes V^a \pd_a ]
\no \\ & \qquad- H_3 [T \otimes V^a \pd_a  \otimes M]  +H_3 [T \otimes V^a \otimes M\pd_a] +H_3 [T \otimes M\pd_a \otimes V^a]
\no \\ & \qquad+ H_3 [V^a \otimes T \otimes M\pd_a] +H_3 [V^a \otimes M\pd_a \otimes T]  + H_3 [M\pd_a \otimes V^a \otimes T]
 \no \\ &\qquad+H_3 [M\pd_a \otimes T \otimes V^a]  - H_3 [M \otimes V^a \pd_a  \otimes T]- H_3 [V^a \pd_a  \otimes M \otimes T ] \no \\
 &\qquad- H_3 [V^a \pd_a   \otimes T \otimes M]\bigg\}\,. \label{TV}
\end{align}
From here one must judiciously act through with the derivatives in each term, moving them either to the left or right, to organise the above expression into cyclic combinations $C_n$.  There is no unique way of achieving this, and it is this step in our approach that is by far the most laborious (especially if there are multiple derivatives present).  In this case, with some familiarity, it is relatively easy to see how to proceed.  For example, the first six $H_3$ terms on the right-hand-side of \eqref{TV} will not include $(\pd_a T)$ arguments if their derivatives are moved through to the right, whereas the last six $H_3$ terms will generate such arguments if their derivatives are moved to the right.  We choose to integrate the last six terms by parts and push their derivatives through to the left so that no $(\pd_a T)$ arguments appear anywhere.  This yields
 \begin{align}
K&(s)\Big|_{a_2,T,V} =\frac{\ii}{2}\tr\int\! \dd k\, k^2 \bigg\{C_3 [M \otimes  (\pd_a V^a)  \otimes  T] - C_3 [T \otimes V^a \otimes (\pd_a M) ] \bigg\} \no \\ &+\frac{1}{2}\tr\int\! \dd k\, (k^2)^2 \bigg\{C_4 [M \otimes  (\pd_a  M)\otimes V^a \otimes  T  ]-C_4 [M \otimes T \otimes V^a \otimes (\pd_a  M)]\bigg\}\,,
\end{align}
which, using the identity \eqref{mr}, simplifies to give
\begin{align}
K(s)\Big|_{a_2,T,V} =&\frac{1}{2}h\,\tr\big(M^{-1}T \pd^a(M^{-1} V_a)\big) =\frac{1}{2}h\,\tr\big(\~T (\pd^a \~V_a)\big)\,. \label{a2tv}
\end{align}
If we had chosen to integrate the first six terms on the right-hand-side of \eqref{TV} by parts, whilst moving the derivative on the last six terms to the right, the result would have been an integrated by parts version of \eqref{a2tv}.  Without performing at least some integration by parts not all terms can be arranged into cyclic combinations.  If needed, one can approach this step of the calculation in a fully systemic fashion by splitting every derivative into a combination of two weighted pieces, for example by writing $\pd_a = \r\, \pd_a +(1-\r) \pd_a$, with $\r$ some arbitrary parameter, and then proceed by acting with one of the weighted pieces to the right and the other to the left.  At the end of the calculation one can then solve for any introduced parameters to ensure the result consists only of cyclic combinations $C_n$.

Overall the final result of using this procedure is\footnote{Recall that we define $\~P :=M^{-1}P$ for any operator $P$, and here $\~V^2 =\~V^a \~V_a$.}
\begin{align}
a_2 &= \tr \bigg(\frac12 \~T^2 -\frac14 \~T \~V^2 +\frac{1}{48}\~V^2\~V^2 +\frac{1}{96}\~V^a\~V^b\~V_a\~V_b - \frac12 \~T (\pd^a\~V_a) -\frac16 \~V^a\~V^b (\pd_b\~V_a) \no \\  & \qquad\qquad\qquad -\frac{1}{12} \~V^b\~V^a (\pd_b\~V_a) +\frac{1}{24}(\pd^a\~V^b)(\pd_a\~V_b)+\frac{1}{12}(\pd^a\~V_a)(\pd^b\~V_b) \bigg)\,. \label{a2b}
\end{align}
Notice that this result is manifestly invariant under the mapping: $M\rightarrow \Upsilon \,M$\,, $T \rightarrow  \Upsilon\, T$ and $V^a \rightarrow  \Upsilon\, V^a$\, for any positive definite spacetime dependent matrix $\Upsilon$.  In particular it demonstrates, by choosing $\Upsilon = M^{-1}$, that up to integration by parts, $a_2$ could equally well have been computed by starting with the \emph{minimal} operator
 \begin{align}
\~\op = M^{-1}\op  = \Box +M^{-1}V^a \pd _a +M^{-1}T\,,
\end{align}
as one would expect from the point of view that $a_2$ originates from an effective action and a field redefinition in the path integral had been invoked.  A general proof of this observation is provided in appendix \ref{proof3}.

It remains an open question as to whether this procedure will generalise to higher order coefficients, though we speculate that it does not. Our attempts in this direction, and the work of \cite{Iochum:2016ynh,Iochum:2017ver}, leads us to conjecture that all higher order coefficients cannot be expressed simply in a closed form in terms of $M$.  We do not anticipate any difficulties in extending these ideas to curved spacetime.

\section{An application: the non-supersymmetric case} \label{sec:app}

As an application of this method we can now quite readily compute the logarithmically divergent contribution to the effective action associated with the following classical action, the non-supersymmetric flat spacetime version of \eqref{ca}, which was studied in curved spacetime in \cite{GKP}:
\begin{align}
S_\mathrm{nss}[A;\f,\ba{\f}]&= -\frac{1}{4} \int\! \dd^4 x\, \,
\Big\{\left(F^{ab}\right)^\mathrm{T} M F_{ab}
	+\left(F^{ab}\right)^\mathrm{T} \L \hat{F}_{ab}\Big\} \no \\&=   \frac{\ii}{2} \int\! \dd^4 x \left(F^{\a\b}\right)^\mathrm{T} \f F_{\a\b} + {\rm c.c.}\,,  \label{nssa}
 \end{align}
where here and below all matrix indices are suppressed.  The above model describes $n$ Abelian gauge fields $A^{a}=\left(A_i^{a}\right)$, $i=1,\ldots,n$, coupled to a complex background field $\f=\f^\mathrm{T}=(\f^{ij})$, where we have introduced the real symmetric $n \times n$ matrices $M$ and $\L$ defined via
\begin{align}
  \f = \L + \ii \,M \,,\label{fdef}
\end{align}
with $M$ positive definite, and $\hat{F}^{ab}= \frac12 \ce^{abcd} F_{cd}$ is the Hodge dual of the electromagnetic field strength $F_{ab}= 2\pd_{[a} A_{b]}$. The second form of the action \eqref{nssa} has been expressed in two-component spinor notation.

The total gauged-fixed action for this model is given by (for details see \cite{GKP})
\begin{equation}
	S_{\mathrm{nss}}^{\mathrm{tot}}[A;\f,\ba{\f}]=\frac{1}{2}\int\! \dd^4x\,\,A_{a}\,\D^a_{\ph{a}b}A^b\,,
\end{equation}
where the operator $\D^a_{\ph{a}b}$ is
\begin{align}
  \D^a_{\ph{a}b} = \d^a_{\ph{a}b} M \Box + V^{a c}_{\ph{ac\,\,}b} \pd _c\,,
\end{align}
where
\begin{align}
V^{acb}=(\pd^c M)\h^{a b}-(\pd^b M)\h^{a c}+(\pd^a M)\h^{c b}-(\pd_d \L)\ce^{acbd}\,. \label{defV}
\end{align}
This operator is only a slight generalisation of the operator \eqref{op} studied in section \ref{sec:Minkowskispace}, since now there is an additional identity matrix carrying spacetime indices in the coefficient of $\Box$.  It is not difficult to see -- by repeating the steps in section \ref{sec:Minkowskispace} -- that in this more general case you arrive at the following expression for the $a_2$ coefficient, which for this model we denote as $a_2^\mathrm{nss}$ (compare with \eqref{a2b}):
\begin{align}
a_2^\mathrm{nss} &= \tr \bigg(\frac{1}{48}\~V^e_{\ph{e}af} \~V^{fa}_{\ph{fa}g}\~V^g_{\ph{g}bh} \~V^{hb}_{\ph{hb}e}
+\frac{1}{96}\~V^{ea}_{\ph{ea}f}\~V^{fb}_{\ph{fb}g}\~V^g_{\ph{a}ah} \~V^h_{\ph{h}be}
 -\frac16 \~V^{ea}_{\ph{ea}f}\~V^{fb}_{\ph{fb}g}(\pd_b \~V^{g}_{\ph{g}ae}) \no \\  & \qquad\qquad
 -\frac{1}{12} \~V^{eb}_{\ph{eb}f}\~V^{fa}_{\ph{fa}g}(\pd_b \~V^{g}_{\ph{g}ae})
 +\frac{1}{24}(\pd^a \~V^{eb}_{\ph{eb}f})(\pd_a \~V^{f}_{\ph{f}be})
 +\frac{1}{12}(\pd^a \~V^{e}_{\ph{e}af})(\pd^b \~V^{f}_{\ph{f}be}) \bigg)\,,
\end{align}
where, as always, we have defined $\~V^{acb} = M^{-1} V^{acb}$. Using the definitions \eqref{fdef} and \eqref{defV}, and simplifying the result (including integration by parts), we arrive at the following:
\begin{align}
	a_2^\mathrm{nss}&=\frac{1}{4}\tr\Big(M^{-1}(\cD^2\f) M^{-1}(\cD^2\ba{\f})\Big)+\frac{1}{24}\tr\Big(M^{-1}(\pd^a\f)M^{-1}(\pd_a\ba{\f})M^{-1}(\pd^b\f)M^{-1}(\pd_b\ba{\f})\Big)\no\\
	&\qquad\qquad +\frac{1}{48}\tr\Big(M^{-1}(\pd^a\f)M^{-1}(\pd^b\ba{\f})M^{-1}(\pd_a\f)M^{-1}(\pd_b\ba{\f})\Big)\,, \label{a2nss}
\end{align}
where we have defined
\begin{align}
	\cD^2\f&:=\Box\f+\ri (\pd^a\f)M^{-1}(\pd_a\f)~,\qquad \cD^2\ba{\f}:=\Box\ba{\f}-\ri (\pd^a\ba{\f})M^{-1}(\pd_a\ba{\f})~.
\end{align}
In restricting the result obtained in \cite{GKP} to flat spacetime, we find agreement with our expression \eqref{a2nss}.  In our case there are no ghost contributions, so $a_2^\mathrm{nss}$ is the only contribution the logarithmically divergent part of the effective action.

\section{Heat kernel calculations in superspace}\label{sec:superspace}

Here we adapt our method and compute the trace of the DeWitt coefficients up to $a_2$ for heat kernels associated with non-minimal operators  in flat $\cN=1$ superspace of the general form
\begin{align}
\bm{\op}&=  \O^{AB} D_A D_B + \O^{A} D_A + \O\,,
\end{align}
where uppercase Latin indices run over spacetime, dotted and undotted, for example $A=(a,\a,\ad)$, and the derivatives $D_A = (\pd_a, D_\a, \DB^{\.{\a}})$. The factors  $\O^{AB}$, $\O^{A}$ and $\O$ are $n\times n$ matrix valued superfields with $\O^{ab} = \h^{ab} M$ and $M=M^\textrm{T}$ positive definite.  This operator has the form $\bm{\op}=  M \Box + \cdots$, and is a generalisation of the vector operator $\op_\mathrm{v}$ defined in \eqref{vectoroperator}.

The heat kernel associated with $\bm{\op}$ is defined by
\begin{align}
  \bm{K}(z,z';s) =\ex^{\ii s \bm{\op}} {\mathbbm 1}_n \d^{(4|4)}(z,z')\,.
\end{align}
In superspace it is convenient to use the supersymmetric interval
\begin{align}
 \label{ip} \z^A = \left\{
 \begin{aligned}
   \,\zeta^a&= x^a-x'^{a}-\ii \q \s_a \ba{\q}'+\ii \q' \s_a \ba{\q}\,,\\\
\zeta^\a&=(\q-\q')^\a\,,\\
\ba{\zeta}_{\.{\a}}&=(\ba{\q}-\ba{\q}')_{\.{\a}}\,,\\
   \end{aligned} \right.
\end{align}
and express the delta function via the integral representation as
\begin{gather}
\d^{(4|4)}(z,z')=\int\!\dd \h\, \ex^{\ii k_a \z^a}\,\ex^{\ii
\l^\a\z_\a}\ex^{\ii
\ba{\l}_{\.{\a}}\ba{\z}^{\.{\a}}}\,, \qquad\qquad \dd \h := 16 \frac{\dd^{4}k}{(2
\p)^4}\dd^{2}\l \dd{^2}\ba{\l}\,.
\end{gather}
One then finds that $\bm{K}(z,z';s)$ becomes
\begin{equation}\label{eq:originalkernel}
\bm{K}(z,z';s)=\int\!\dd \h \;\ex^{\ii k^a \z_a} \ex^{\ii
\l^\a \z_\a} \ex^{\ii\ba{\l}_{\.{\a}}\ba{\z}^{\.{\a}}} \ex^{ \ii s \hat{\bm{\op}}}\,,
\end{equation}
where
\begin{align}
\hat{\bm{\op}} = \O^{AB} X_A X_B + \O^{A} X_A + \O\,,
\end{align}
with the $X$'s being shifted covariant derivatives defined by
\begin{align}
  X_A = \left\{
 \begin{aligned}
   \, X_a&=\pd_a+\ii k_a\,, \\
X_{\a}&=D_{\a}+\ii\l_\a-k_{\a\.{\a}}\ba{\z}^{\.{\a}}\,,
\\ \ba{X}^{\.{\a}}&=\DB^{\.{\a}}+\ii\ba{\l}^{\.{\a}}
-k^{\a\.{\a}}\z_{\a}\,.\\
   \end{aligned} \right.
\end{align}

Again we are interested in the functional trace of the heat kernel, which is given by
\begin{align}
\bm{K}(s) = \int\! \dd^{4|4} z\,  \bm{K}(z;s)\,,
\end{align}
where
\begin{align}
\bm{K}(z;s) = \lim_{z' \rightarrow z} \tr\, \bm{K}(z,z';s) = \tr \int\! \dd \h\, \ex^{\ii s \hat{\bm{\op}} }\,,  \label{k}
\end{align}
which has the asymptotic expansion
\begin{align}
\bm{K}(z;s) = \frac{h}{s^2}\sum_{n=0}^{\infty}(\ii s)^n\,  \bm{a}_n(z)\,,
\end{align}
with $\bm{a}_n(z)$ the trace of the heat kernel coefficients in the coincidence limit.

Expanding the operator $\hat{\bm{\op}}$ in a power series of the integral parameters $k^A=(k^a,\l^\a, \ba{\l}_{\.\a})$,  we find
\begin{align}
\hat{\bm{\op}} = \bm{\op} +k_A \Psi^A+k_A k_B \Psi^{AB}\,,
\end{align}
where the objects $\Psi^A$ are first order differential operators, whilst the $\Psi^{AB}$ are just matrices which come solely from $\O^{AB}$.  Rescaling the integral parameters $k^A \rightarrow k^A/\sqrt{s}$, under which the integration measure $\dd \h$ is unchanged, we find
\begin{align}
\bm{K}(z;s)= \tr\int\! \dd \h\, \exp\Big\{\ii s \bm{\op} +\ii\sqrt{s} k_A \Psi^A+ \ii k_A k_B \Psi^{AB}\Big\} \,. \label{hk}
\end{align}
This clearly has a power series expansion in $s$ beginning with order $s^0$, and so the heat kernel coefficients $\bm{a}_0$ and $\bm{a}_1$ vanish, a consequence of supersymmetry.  To compute $\bm{a}_2$ we can now freely set $s=0$ in \eqref{hk} leaving
\begin{align}
\bm{a}_2 = -\frac{1}{h} \tr\int\! \dd \h\, \exp\Big\{\ii k_A k_B \Psi^{AB}\Big\}\,.
\end{align}
Since this no longer contains any differential operators we can also freely set $z'= z$ in $\Psi^{AB}$.  From this we can see that  $\O^{A}$ and $\O$ do not contribute to $\bm{a}_2$ and so may be ignored.  For explicit computation we now write the operator $\bm{\op}$ as
\begin{align}
\bm{\op} &= M \Box +U D^2 + Q\DB^2 +P^{a\a}\pd_a D_{\a}+N^{a\ad}\pd_a \DB_{\ad}+H^{\a\.{\a}}[D_\a\,,\DB_{\.{\a}}] +\cdots \label{op2}
\end{align}
where the ellipsis indicates the irrelevant $\O^{A}$ and $\O$ pieces.  This yields
\begin{align}
\bm{a}_2 = -\frac{1}{h}\tr\int\! \dd \h\,\ex^{A+B}\,, \label{hka}
\end{align}
with
\begin{align}
  A =- \ii k^2 M\,,\qquad
 B = \ii k_a \l_\a P^{a\a} + \ii k_a \ba{\l}_{\.{\a}} N^{a{\.{\a}}} - 2\ii \l_\a \ba{\l}_{\.\a}H^{\a \.{\a}}-  \ii \l^2 U -  \ii \ba{\l}^2 Q\,.
\end{align}
We now expand \eqref{hka} using the Dyson series:
\begin{align}
\bm{a}_2  = -\frac{1}{h} \tr \int\! \dd \h \,\bigg\{\ex^A + \sum_{n=1}^{\infty} H_n [B \otimes B \otimes \cdots \otimes B]\bigg\}\,. \label{hke2}
\end{align}
Due to the integral over the fermionic parameters $\l$ and $\ba{\l}$, only terms in this expansion of order $\l^2 \ba{\l}^2$ will survive, leaving us with
\begin{align}
\bm{a}_2  = -\frac{1}{h} \tr\int\! \dd \h \,\bigg\{H_2 [B \otimes B]+  H_3 [B \otimes B \otimes B]+ H_4 [B \otimes B \otimes B \otimes B]\bigg\}\,.
\end{align}
Expanding out the factors of $B$ and retaining only terms proportional to $\l^2 \ba{\l}^2$, we find
\begin{align}
\bm{a}_2 & = -\frac{16}{h} \tr\int\! \dd k \,\bigg\{-\frac12 C_2 [H^{\a\.{\a}}\otimes H_{\a\.{\a}}] - C_2 [Q\otimes U] +\frac{\ii k^2}{8}\Big(C_3 [ N^{a \.{\a}}\otimes N_{a \.{\a}} \otimes V] \no \\ &+C_3 [ N^{a \.{\a}}\otimes P_a^{\phantom{a}\a} \otimes H_{\a\.{\a}}] -C_3 [ N^{a \.{\a}}\otimes H^{\a}_{\phantom{\a}\.{\a}} \otimes P_{a\a}]-C_3 [ Q\otimes P^{a\a} \otimes P_{a\a}]\Big) \no \\&+\frac{R_{abcd}(k^2)^2}{192}\Big(C_4[ N^{a \.{\a}}\otimes  P^{b\a} \otimes  N^{c}_{\phantom{c}\.{\a}}\otimes P^d_{\phantom{d}\a}] -2 C_4[ N^{a \.{\a}}\otimes N^{b}_{\phantom{c}\.{\a}}   \otimes P^{c\a}  \otimes P^d_{\phantom{d}\a}]\Big)\bigg\}\,,
\end{align}
where, in this more general context, cyclic combinations $C_n$ take into account the parity $\ce(B_i)$ of the matrices $B_i$,
\begin{align}
  C_n [B_1 \otimes B_2 & \otimes \cdots  \otimes B_n] :=  H_n [B_1 \otimes B_2 \otimes \cdots \otimes B_n] \no \\ \qquad & +(-1)^{\ce(B_n)} H_n [B_n \otimes B_1 \otimes B_2 \otimes \cdots  \otimes B_{n-1}] \no \\ &\qquad+ (-1)^{\ce(B_n)+\ce(B_{n-1})}H_n [B_{n-1} \otimes B_n \otimes B_1 \otimes \cdots  \otimes B_{n-2}] +  \cdots \no \\ & \qquad\qquad+  (-1)^{\ce(B_n)+\ce(B_{n-1})+\cdots +\ce(B_2)}H_n [B_2 \otimes B_3 \otimes \cdots \otimes B_n \otimes B_1]\,, \label{cp}
\end{align}
where the central identity \eqref{mr} still holds (the proof is trivially extended to this more general setting).

Our final result is
\begin{align}
\bm{a}_2  &= \tr \bigg( 8 \~H^{\a\.{\a}}\~H_{\a\.{\a}} + 16 \~Q \~U - 2 \~N^{a \.{\a}}\~N_{a \.{\a}}\~U-2 \~N^{a \.{\a}}\~P_{a}^{\phantom{a} \a}\~H_{\a\.{\a}}
+2 \~N^{a \.{\a}}\~H^{\a}_{\phantom{\a} \.{\a}}\~P_{a\a}+ 2 \~Q\~P^{a\a}\~P_{a\a} \no \\ & \quad\quad\quad
- \frac16  \~N^{a \.{\a}}\~N_{a \.{\a}}\~P^{b\a}\~P_{b\a}
- \frac16 \~N^{a \.{\a}}\~N^b_{\phantom{b} \.{\a}}\~P_{b}^{\phantom{b}\a}\~P_{a\a}
- \frac16  \~N^{a \.{\a}}\~N^b_{\phantom{b} \.{\a}}\~P_{a}^{\phantom{a}\a}\~P_{b\a} \no \\
&\qquad\qquad
+ \frac{1}{12}  \~N^{a \.{\a}}\~P_a^{\phantom{a} \a}\~N^{b}_{\phantom{b}\.{\a}}\~P_{b\a} + \frac{1}{12}  \~N^{a \.{\a}}\~P^{b \a}\~N_{b\.{\a}}\~P_{a\a}
+ \frac{1}{12}  \~N^{a \.{\a}}\~P^{b \a}\~N_{a\.{\a}}\~P_{b\a}
\bigg)\,. \label{a2susy}
\end{align}
As with the case in Minkowski space, we once again note that the above result is manifestly inert under the following mapping:
\begin{subequations}
\begin{gather}
 M\rightarrow \Upsilon\, M\,,\quad\quad U \rightarrow  \Upsilon\, U\,,\quad\quad Q \rightarrow  \Upsilon\, Q,\quad\quad P^{a\a} \rightarrow  \Upsilon\, P^{a\a}\,, \\  N^{a\.{\a}} \rightarrow  \Upsilon\, N^{a\.{\a}},\quad\quad H^{\a\.{\a}} \rightarrow  \Upsilon\, H^{\a\.{\a}}\,,
\end{gather}
\end{subequations}
where $\Upsilon$ is any positive definite matrix.  This demonstrates, by choosing $\Upsilon = M^{-1}$, that up to integration by parts, $\bm{a}_2$ could equally well have been computed by using the minimal operator $\~{\bm{\op}} = M^{-1}\bm{\op}$.

If we now specialise to the vector multiplet operator of interest, $\bm{\op}=\op_\mathrm{v}$, achieved by setting
\begin{subequations}
\begin{gather}
  M =\X\,, \qquad N^{a \.{\a}} = - \frac{\ii }{2} (\s_a)^{\a\.{\a}} (D_{\a}\X)\,,\qquad P^{a {\a}} =- \frac{\ii }{2} (\s_a)^{\a\.{\a}} (\DB_{\.{\a}}\X)\,, \\ Q=\frac{1}{16}(D^2 \X)\,, \qquad  U=\frac{1}{16}(\DB^2 \X)\,, \qquad H^{\a\.{\a}} = 0\,,
\end{gather}
\end{subequations}
we find that \eqref{a2susy} simplifies to give the  corresponding heat kernel coefficient
\begin{align}
 a_{2}^\mathrm{v}  = \tr \bigg(\frac{1}{16}\X^{-1}(\nabla^2\X)\X^{-1}(\ba{\nabla}^2\X) -\frac{1}{8}\X^{-1}(D^\a \X)\X^{-1}(\DB_{\.{\a}} \X)\X^{-1}(D_\a \X)\X^{-1}(\DB^{\.{\a}} \X)\bigg) \label{a2v}
\end{align}
where we have defined
\begin{align}
  \nabla^2\X := D^2 \X - 2  (D^\a \X)\X^{-1}(D_\a \X)\,, \qquad
   \ba{\nabla}^2\X := D^2 \X - 2  (\DB_{\.{\a}} \X)\X^{-1}(\DB^{\.{\a}} \X)\,.
\end{align}
This is consistent with the result found in \cite{K2020}.

\section{Heat kernel calculations for chiral operators}\label{sec:chiral}

Here we apply our method and compute the trace of the DeWitt coefficients up to $a_2$ for heat kernels associated with non-minimal chiral operators  in flat $\cN=1$ superspace of the general form
\begin{align}
\bm{\op}_c&=  M \Box + L^{a\a }\pd_a D_\a + N D^2 +F^{\a} D_\a + J^a \pd_a +G\,,
\end{align}
where all of the coefficients of the derivatives, $M$, $L^{a\a },\ldots,$ are understood as being $n\times n$ matrix valued superfields with $M=M^\textrm{T}$ positive definite.  Additionally, we also require $\DB_{\.{\a}} \bm{\op}_c \f = 0$ for chiral scalars $\f = (\f_i)$, $\DB_{\.{\a}} \f =0$, which implies the following constraints on the coefficients:
\begin{subequations}\label{cc}
\begin{gather}
\DB_{\.{\a}}N=0\,, \qquad\DB_{\.{\a}}F_\a =0\,,\qquad \DB_{\.{\a}}G =0 \,,\qquad \DB_{\.{\a}}J_a +2 \ii (\s_{a})_{\a\.{\a}}F^\a=0\,,\\
 \DB_{\.{\a}}L_{a\a} - 4 \ii (\s_a)_{\a\.{\a}} N =0\,,\qquad  \h_{ab}\DB_{\.{\a}}M +2 \ii (\s_{(a})_{\a\.{\a}} L_{b)}^{\phantom{b}\a}=0\,.
\end{gather}
\end{subequations}

The associated chiral heat kernel is
\begin{align}
  \bm{K}_c(z,z';s) =\ex^{\ii s \bm{\op}_c} {\mathbbm 1}_n \d_{+}(z,z')\,.
\end{align}
Using the definitions of $\z^a$ and $\z^\a$ introduced in \eqref{ip}, we express the chiral delta function $\d_{+}(z,z')$ via the integral representation as
\begin{gather}
\d_{+}(z,z')=\int\!\dd \h_c\, \ex^{\ii k_a \z^a}\,\ex^{\ii
\l^\a\z_\a}\,, \qquad\qquad \dd \h_c: = 4 \frac{\dd^{4}k}{(2
\p)^4}\dd^{2}\l\,,
\end{gather}
and  find that the trace of the chiral heat kernel in the coincidence limit is
\begin{equation}
\bm{K}_c(z;s) = \lim_{z' \rightarrow z} \tr\, \bm{K}_c(z,z';s)=\int\!\dd \h_c \,\ex^{ \ii s \hat{\bm{\op}}_c}\,,
\end{equation}
where
\begin{align}
\hat{\bm{\op}}_c =  M X^a X_a + L^{a\a} X_a X_\a + N X^2 +F^{\a} X_\a + J^a X_a +G\,.  \label{opc}
\end{align}
Here the $X$'s are shifted derivatives defined by\footnote{Note that there is also a shift of $-k_{\a\.{\a}}\ba{\z}^{\.{\a}}$ in the derivative $D_\a$, however this contribution always vanishes in the coincidence limit since there are no $\DB_{\.\a}$ operators present.}
\begin{align}
X_a=\pd_a+\ii k_a\,, \qquad \qquad X_{\a}=D_{\a}+\ii\l_\a.
\end{align}

Again we are interested in the functional trace of the chiral heat kernel, which is given by
\begin{align}
\bm{K}_c(s) = \int\! \dd^4 x \dd^2 \q\,  \bm{K}_c(z;s)\,,
\end{align}
with asymptotic expansion
\begin{align}
\bm{K}_c(z;s) = \frac{h}{s^2}\sum_{n=0}^{\infty}(\ii s)^n\,  \bm{a}^c_n(z)\,,
\end{align}
with $\bm{a}^c_n(z)$ the trace of the chiral heat kernel coefficients in the coincidence limit.

Expanding the operator \eqref{opc} in a power series of the integral parameters $k^a$ and $\l^\a$ we write
\begin{align}
\hat{\bm{\op}}_c = \bm{\op}_c +k^a Y_a+ \l^\a Z_\a +\l_\a k_a L^{a\l}-\l^2 N - k^2 M\,,
\end{align}
with
\begin{align}
  Y_a = 2 \ii M \pd_a + \ii L_{a}^{\phantom{a}\a} D_\a + \ii J_a\,, \qquad  Z_\a  = \ii L^{\phantom{\a}a}_{\a}\pd_a +2 \ii N D_\a + \ii F_\a \,.
\end{align}
We now rescale $k^a \rightarrow k^a/\sqrt{s}$ and $\l^\a \rightarrow \l^\a/\sqrt{s}$, under which the integration measure $\dd \h_c \rightarrow \dd \h_c/s$, giving
\begin{align}
\bm{K}_c(z;s) = \frac{\tr}{s}\int\! \dd \h_c\, \ex^{A+B}\,,
\end{align}
with
\begin{align}
 A= - \ii k^2 M\,, \qquad  B= \ii s \bm{\op}_c +\ii\sqrt{s} k^a Y_a+ \ii\sqrt{s}\l^\a Z_\a
+\ii \l_\a k_a L^{a\l}-\ii \l^2 N\,.
\end{align}
Expanding the kernel $\bm{K}_c(z;s)$ using the Dyson series, we can now identity expressions for the first three DeWitt coefficients (keeping in  mind that only terms proportional to $\l^2$ will contribute).  The first coefficient trivially vanishes, $\bm{a}_0^c =0$.  The second coefficient is found to be
 \begin{align}
\bm{a}_1^c &=-\frac{\ii}{h} \tr \int\! \dd k\,\Big\{- 4 \ii H_1[N] + \frac{k^2}{2} H_2[L^{a\a}\otimes L_{a\a}]\Big\} \no \\
&=-\tr \Big(4 M^{-1} N M^{-1}+\frac12 M^{-1} L^{a\a} M^{-1} L_{a\a}M^{-1} \Big) \,,
 \end{align}
where in the last line we have used manipulations similar to \eqref{a0msa} and \eqref{a0msb}.

Computing the third coefficient, $\bm{a}_2^c$, requires a substantial amount of work, but at this stage the procedure is routine.  We first isolate the $\bm{a}_2^c$ coefficient in the Dyson series, which we find to be:
\begin{multline}
\bm{a}_2^c = -4\frac{\tr}{h} \int\! \dd k \Bigg\{H_2[\bm{\op}_c \otimes N] +H_2[N\otimes \bm{\op}_c ]+ \frac12 H_2[Z^\a\otimes Z_\a]  \\
+\frac{\ii  k^2}{8}\Big(H_3[\bm{\op}_c \otimes L^{a\a} \otimes L_{a\a} ]+H_3[L^{a\a} \otimes \bm{\op}_c \otimes L_{a\a} ] +H_3[L^{a\a} \otimes L_{a\a} \otimes  \bm{\op}_c]  \\+2H_3[Y^a \otimes Y_a \otimes N ]+2H_3[Y^a \otimes N \otimes Y_a ]+2H_3[N\otimes Y^a  \otimes Y_a ] \\
-H_3[Y^a \otimes Z^\a \otimes L_{a\a} ]
-H_3[Y^a \otimes L_a^{\phantom{a}\a}  \otimes Z_\a ]
-H_3[Z^\a \otimes Y^a  \otimes  L_{a\a} ] \\
-H_3[Z^\a \otimes L^a_{\phantom{a}\a} \otimes Y_a  ]
-H_3[L^{a\a}  \otimes Y_a   \otimes Z_\a]
-H_3[L^{a\a}  \otimes Z_\a  \otimes Y_a]\Big) \\
-\frac{(k^2)^2}{48}R_{abcd}\Big(
H_4[Y^{a}  \otimes Y^{b}  \otimes L^{c\a}\otimes L^{d}_{\phantom{d}\a}]+
H_4[Y^{a}  \otimes L^{b\a}  \otimes Y^{c}\otimes L^{d}_{\phantom{d}\a}]+
H_4[Y^{a}  \otimes L^{b\a}  \otimes L^{c}_{\phantom{c}\a}\otimes Y^{d}] \\+
H_4[L^{a\a}  \otimes Y^{b}  \otimes Y^{c} \otimes L^{d}_{\phantom{d}\a}]+
H_4[L^{a\a}  \otimes Y^{b} \otimes L^{c}_{\phantom{d}\a} \otimes Y^{d} ]+
H_4[L^{a\a}  \otimes L^{b}_{\phantom{d}\a} \otimes Y^{c} \otimes Y^{d} ]\Big)\Bigg\}\,.
\end{multline}
As before, we now use Gaussian moment-like generating identities, and act with the derivatives $\pd_a$ and $D_\a$ either left or right in such a way as to organise all terms into cyclic combinations $C_n$ of the form \eqref{cp}.  Then using the identity \eqref{mr} we compute the final result, which is quite long and for completeness is displayed in appendix \ref{appB}, equation \eqref{ca2appB}.

Similar to the cases of Minkowski and full superspace, we note the surprising outcome that the final result \eqref{ca2appB} is manifestly inert under the mapping:
\begin{subequations}
\begin{gather}
 M\rightarrow \Upsilon\, M\,,\quad\quad L_{a\a} \rightarrow  \Upsilon\, L_{a\a}\,,\quad\quad N \rightarrow \Upsilon\, N,\quad\quad  F^{\a} \rightarrow  \Upsilon\, F^{\a}\,,\\
 J^a \rightarrow  \Upsilon\, J^a\,, \quad\quad G \rightarrow  \Upsilon\, G\,,
\end{gather}
\end{subequations}
for any positive definite matrix $\Upsilon$.  This demonstrates,  by choosing $\Upsilon = M^{-1}$, that up to integration by parts, $\bm{a}_2^c$ could at least formally have been computed by starting with a \emph{non-chiral} minimal operator $\~{\bm{\op}}_c = M^{-1}\bm{\op}_c$.

If we now specialise to the chiral operator of interest, $\bm{\op}_c=\op_{+}$, by setting
\begin{gather}
  M =\X^2\,, \qquad L_{a \a} =  \frac{\ii }{2} (\s_a)_{\a\.{\a}} (\DB^{\.{\a}}\X^2)\,,\qquad N=\frac{1}{16}(\DB^2 \X^2)\,,\\
  J_{a } =- \frac{\ii }{2} (\s_a)_{\a\.{\a}} \DB^{\.{\a}}(\X D^\a \X)\,,\qquad F^\a = \frac18 \DB^2(\X D^\a \X)\,, \qquad  G=\frac{1}{16}\DB^2( \X D^2 \X)\,,
\end{gather}
we find, after some simplification, the corresponding heat kernel coefficient, $a_2^{+}$, can be cast on full superspace and the result is
\begin{multline}
  \int\! \dd^4 x \dd^2 \q \,a_2^{+} = \frac{\tr}{96}\int\! \dd^{4|4} z \,\Big\{-2 e_{0,1,2}-2 e_{0,2,1}+2 e_{1,0,2}+2 e_{2,0,1}+2 f_{0,1,2}+2 f_{0,2,1}\\ \qquad-2 f_{1,0,2}-2   f_{2,0,1}-g_{1,0,1,2}-g_{1,0,2,1}+g_{1,1,2,0}+g_{1,2,1,0}-g_{2,0,1,1}+g_{2,1,1,0}-h_{0,2,0,2}\\ \qquad-h_{1,1,0,2} -6
   h_{1,1,1,1}+h_{1,1,2,0}-h_{1,2,0,1}+h_{1,2,1,0}+h_{2,0,2,0}-h_{2,1,0,1}+h_{2,1,1,0}\Big\}\,,
\end{multline}
where we have made the following identifications for positive integers $p,q,r,s$:
\begin{subequations}
\begin{align}
e_{p,q,r}&=\tr\big((\DB^{\.{\a}}D^\a \X)\X^{-p} (\DB_{\.{\a}} \X) \X^{-q}(D_\a \X)\X^{-r}\big)\,,\\
f_{p,q,r}&=\tr\big((D^\a \DB^{\.{\a}}\X)\X^{-p} (D_\a \X) \X^{-q}(\DB_{\.{\a}} \X)\X^{-r}\big)\,,\\
g_{p,q,r,s}&=\tr\big((D^\a \X)\X^{-p} (D_\a \X) \X^{-q}(\DB_{\.{\a}} \X)\X^{-r} (\DB^{\.{\a}} \X) \X^{-s}\big)\,,\\
h_{p,q,r,s}&=\tr\big((D^\a \X)\X^{-p} (\DB_{\.{\a}} \X) \X^{-q}(D_\a \X)\X^{-r} (\DB^{\.{\a}} \X) \X^{-s}\big)\,.
\end{align}
\end{subequations}

Finally, adding this result to its complex conjugate, we find arrive at the final outcome
\begin{align}
  \int\! \dd^4 x \dd^2 \q \,a_2^{+} + \mathrm{c.c.}= -\frac18 \tr \int\! \dd^{4|4} z \,\X^{-1}(D^\a \X)\X^{-1} (\DB_{\.{\a}} \X) \X^{-1}(D_\a \X)\X^{-1} (\DB^{\.{\a}} \X)\,, \label{fca2}
\end{align}
in agreement with the results of \cite{K2020} for the $n=1$ case.

\section{Final results and generalisations}\label{sec:final}

In terms of the functional trace of the heat kernels $K_\mathrm{v}(s)$ and $K_\mathrm{\pm}(s)$ associated with the operators $\op_\mathrm{v}$ and $\op_\mathrm{\pm}$ respectively,  the regularised effective action \eqref{ee3} is
\begin{align}
  \G [\F, \bar \F] = \frac{\m^{2 \w}}{2}\int_{0}^{\infty} \frac{\dd s}{ (\ii s)^{1-\w}}\Big\{K_\mathrm{v}(s)- \frac12\big(K_{+}(s) +\mathrm{c.c.}\big)\Big\}\,,
\end{align}
which leads to the following expression for its logarithmically divergent contribution:
\begin{align}
  \G_\mathrm{div}[\F, \bar \F] &= \frac{1}{32 \p^2 \w}\Bigg\{\int \! \dd^{4|4} z\, a_2^\mathrm{v}- \frac12\left(\int\!  \dd^4 x \dd^2 \q \,a_2^{+} +\mathrm{c.c.}\right)\Bigg\}\,.
\end{align}
Inserting the expressions \eqref{a2v} and \eqref{fca2}, our final overall result is
\begin{align}
  \G_\mathrm{div}[\F, \bar \F] &= \frac{1}{512 \p^2 \w} \int \! \dd^{4|4} z\,{\rm tr}\, \Big\{ \X^{-1}(\nabla^2\X)\X^{-1}(\ba{\nabla}^2\X) \qquad\qquad\no \\ &\qquad\qquad\qquad\qquad\qquad -\X^{-1}(D^\a \X)\X^{-1}(\DB_{\.{\a}} \X)\X^{-1}(D_\a \X)\X^{-1}(\DB^{\.{\a}} \X) \Big\}~.
 \label{6.3}
\end{align}
This result reduces to that derived in \cite{K2020} in the $n=1$ case.

The functional \eqref{6.3} can be rewritten as a higher-derivative superconformal
$\s$-model on the Hermitian symmetric  space $\sSp(2n,\dsR)/\sU(n)$:
\begin{subequations}
\bea
\G_\mathrm{div}[\F, \bar \F] &=& \frac{1}{128 \p^2 \w} S_{\rm ind}~,
\eea
where $S_{\rm ind}$ denotes the induced action and has the following explicit form:
\bea
S_{\rm ind}&=&
\frac{1}{16} \int
\rd^{4|4}z\,
 \bigg\{
{\mathfrak g}_{I \bar J} (\F, \bar \F)
\nabla^2 \F^I \bar \nabla^2 \bar \F^{\bar J} \non \\
&&\qquad \qquad +\hf {\mathfrak R}_{I \bar J K \bar L} (\F, \bar \F)
 \cD^\a \F^I \cD_\a \F^K \bar \cD_\ad
\bar \F^{\bar J} \bar \cD^\ad \bar \F^{\bar L} \bigg\} ~.
\label{6.4b}
\eea
\end{subequations}
Here we have used the condensed notation $\F^I \equiv \F^{ij}$. The expressions for the K\"ahler metric
${\mathfrak g}_{I \bar J} $ and the Riemann tensor ${\mathfrak R}_{I \bar J K \bar L} $
on $\sSp(2n,\dsR)/\sU(n)$ are explicitly given, for example, in \cite{GKP}.
This action has a unique super-Weyl invariant extension to curved superspace of the
form \eqref{sigma} with
\bea
{\mathfrak F}_{IJ \bar K \bar L} (\F, \bar \F)
= \hf {\mathfrak R}_{I \bar K J \bar L} (\F, \bar \F)~.
\eea

 The model \eqref{ca} has a natural extension to the case of local $\cN=2$ supersymmetry \cite{K2020} that describes the coupling of
 $n$ Abelian ${\cal N}=2$ vector multiplets to ${\cal N}=2$ chiral multiplets
  parametrising $\mathsf{Sp}(2n, {\mathbb R})/ \mathsf{U}(n)$. The action is
 \bea
S_{\cN=2}  [{\mathbb V}; X , \bar X]= - \frac{\ri }{8} \int \rd^4x \rd^4 \q  \, \cE \,
{\mathbb W}_i X^{ij} {\mathbb W}_j
+{\rm c.c.}
\label{VM-actionN=2}
\eea
Here ${\mathbb V}_i$ denotes vector multiplet prepotentials (which we do not specify in this paper),
$X^{ij} $ are  background chiral scalar superfields, ${\bar \cD}^{\ad }_{\hat i} X^{ij}=0$,
and  ${\mathbb W}_i$ are the field strengths of $n$ vector multiplets.
   The latter are reduced chiral superfields,
 \bea
{\bar \cD}^{\ad }_{\hat i} {\mathbb W}_i =0~, \qquad
\Big(\cD^{\hat i \hat j}+4S^{\hat i \hat j}\Big){\mathbb W}_i
=
\Big(\cDB^{\hat i \hat j}+ 4\bar{S}^{\hat i \hat j}\Big)\bar{\mathbb W}_i~.
\label{vectromul}
\eea
Here we have used the notation $\cD^{\hat i \hat j} = \cD^{\a (\hat i}  \cD_\a^{\hat j)}$,
$\bar \cD^{\hat i \hat j} = \bar \cD_\ad ^{ (\hat i}  \bar \cD^{ \ad \hat j)}$,
with $\cD_A = (\cD_a, \cD_\a^{\hat i} , \bar \cD^\ad_{\hat i}) $ the supergravity covariant derivatives \cite{KLRT-M}.
The $\sSU(2)$ indices are denoted $\hat i$, $\hat j$ etc. The tensors
${S}^{\hat i \hat j}$ and $\bar{S}^{\hat i \hat j}$
in \eqref{vectromul}
are special components of the torsion tensor, see \cite{KLRT-M} for the technical details. The model \eqref{VM-actionN=2} is invariant under $\cN=2$ super-Weyl transformations (assuming $X^{ij}$ to be super-Weyl inert)
and possesses  the maximal possible duality group, $\sSp(2n, {\mathbb R})$.\footnote{The concept of duality invariant models for $\cN=2$ vector multiplets was introduced in \cite{KT1,KT2}.}

We are interested in computing the logarithmically divergent part of the effective action for the model \eqref{VM-actionN=2}, which is obtained by integrating out the $\cN=2$ vector multiplets. The simplest way to achieve this consists of two steps: (i) restrict the model \eqref{VM-actionN=2} to $\cN=2$ Minkowski superspace; and (ii) reduce the resulting classical action to $\cN=1$ Minkowski superspace.
The $\cN=2$
chiral superfield $X^I\equiv X^{ij} = X^{ji}$, $\bar D^\ad_i X^I=0$, is equivalent to three $\cN=1$ chiral superfields $\F^I$, $\O^I_\a$ and $Z^I$  defined as follows:
\bea
\F^I :=X^I \big|_{\q_{\hat 2}=0} ~, \qquad
\sqrt{2}  \Omega^I_\a := D^{\hat 2}_\a X^I \big|_{\q_{\hat 2} =0}~,
\qquad Z^I := -\frac 14 (D^{\hat 2})^2 X^I\big|_{\q_{\hat 2}=0} ~.
\eea
As an example of applying $\cN=2 \to \cN=1$ reduction, we consider a higher-derivative $\cN=2$ supersymmetric nonlinear $\s$-model
\bea
S =   \int  \rd^4x \rd^4 \q  \rd^4 \bar \q \, {\mathfrak K}(X, \bar X )
  = \frac{1}{16} \int \rd^4x \rd^2 \q  \rd^2 \bar \q \,
  (D^{\hat 2})^2 (\bar D_{\hat 2})^2
  {\mathfrak K}(X, \bar X )\Big|_{\q_{\hat 2} = \bar \q^{\hat 2}= 0}~.
\eea
Direct calculation now gives \cite{K2020}
\bea
S &=& \frac{1}{16}  \int
\rd^{4|4} z
 \,   \bigg\{ {\mathfrak g}_{I \bar J}
\nabla^2 \F^I \bar \nabla^2 \bar \F^{\bar J}
+ {\mathfrak R}_{I \bar J K  \bar L}   D^\a \F^I D_\a \F^K \bar D_\ad
\bar \F^{\bar I} \bar D^\ad \bar \F^{\bar L} \bigg\}  \non \\
&& + \int
\rd^{4|4} z
\,  {\mathfrak g}_{I \bar J} \bigg\{ {\mathbb Z}^I \bar {\mathbb Z}^{\bar J}
-{\ri} \O^{I \a} \nabla_{\a\ad} \bar \O^{\bar J \ad} \bigg\}~,
\label{6.10}
\eea
where we have defined
\bea
{\mathbb Z}^I = Z^I -\frac 14 \G^I_{JK} \O^{J\a} \O^K_\a ~, \qquad
\nabla_{\a\ad} \bar \O^{\bar I \ad} = \pa_{\a\ad} \bar \O^{\bar J \ad}
+\G^{\bar I}_{\bar J \bar K} \pa_{\a\ad}\bar \F^{\bar J} \bar \O^{\bar K \ad}~.
\eea
The superfield ${\mathbb Z}^I$ transforms as a target-space vector under holomorphic reparametrisations of the K\"ahler manifold,  but it is not chiral unlike $Z^I$.
It is important to point out that
the higher-derivative $\cN=1$ supersymmetric nonlinear $\s$-model
in the first line in \eqref{6.10} is similar to \eqref{6.4b}.

The  $\cN=2$ chiral field strength ${\mathbb W}_i$
contains two independent chiral $\cN=1$ components
\bea
\sqrt{2}  \J_i := {\mathbb W}_i | ~, \qquad 2{\rm i} W_{\a i}:= D_\a^{\hat{2}}\, {\mathbb W}_i |\quad \implies \quad (D^{\hat{2}})^2{\mathbb W}_i|
= \sqrt{2} \bar{D}^2\bar\J_i~.
\eea
Now applying the $\cN=2 \to \cN=1$ reduction to the flat-superspace version of \eqref{VM-actionN=2} gives
\bea
S_{\cN=2} =  -\frac{\ii}{4}\int\!
   \rd^4x \rd^2 \q \,
  \, W_i^\a \F^{ij} W_{\a j} +\mathrm{c.c.}
  + \hf \int \rd^{4|4}z\, \bar \J_i \X^{ij} \J_j +\dots,
  \label{6.13}
\eea
where the symmetric matrix $\X$ is defined by \eqref{111},
and the ellipsis denotes the contributions containing the superfields $\O^{ij}_\a$,
$Z^{ij}$ and their conjugates.

In the path integral, the second term in \eqref{6.13} generates a contribution which cancels out the second and third terms in \eqref{ee3}. Consequently the effective action is given by
\begin{align}
  \G_{\cN=2}  [\F, \bar \F] = \frac{\ii}{2} \Tr \ln \op_\mathrm{v}  ~.
 \end{align}
As a result, the divergent part of the effective action is
\begin{subequations}
\bea
\G^{( \cN=2)}_\mathrm{div}[\F, \bar \F] &=& \frac{1}{128 \p^2 \w} S^{( \cN=2)}_{\rm ind}~,
\eea
where $S^{( \cN=2)}_{\rm ind}$ denotes the induced action and has the following explicit form:
\bea
S^{( \cN=2)}_{\rm ind}&=&
\frac{1}{16} \int
\rd^{4|4}z\,
 \bigg\{
{\mathfrak g}_{I \bar J} (\F, \bar \F)
\nabla^2 \F^I \bar \nabla^2 \bar \F^{\bar J} \non \qquad\qquad\qquad \\
&&\qquad \qquad\qquad + {\mathfrak R}_{I \bar J K \bar L} (\F, \bar \F)
 \cD^\a \F^I \cD_\a \F^K \bar \cD_\ad
\bar \F^{\bar J} \bar \cD^\ad \bar \F^{\bar L} \bigg\} + \dots
\eea
\end{subequations}
We see that  the $\F$-dependent part of the induced action coincides with
the higher-derivative $\cN=1$ supersymmetric nonlinear $\s$-model
in the first line of \eqref{6.10}.
Since the complete induced action must be $\cN=2$ supersymmetric,
we naturally arrive at the following functional
 \bea
S^{( \cN=2)}_{\rm ind} =  -4  \int  \rd^4x \rd^4 \q  \rd^4 \bar \q \, E\,
{\rm tr} \ln \Big[\ri (\bar X - X) \Big]~,
\label{6.16}
\eea
which is $\cN=2$ locally supersymmetric and super-Weyl invariant.
Here ${\mathfrak K}(X, \bar X ) = - 4 \,
{\rm tr} \ln \Big[\ri (\bar X - X) \Big]$ is the K\"ahler potential of  $\sSp(2n,\dsR)/\sU(n)$.

We point out that associated with any K\"ahler manifold ${\mathfrak M}^n$
is a higher-derivative $\cN=2$ superconformal  $\sigma$-model of the form
 \cite{BdeWKL,deWKvZ,GHKSST}
 \bea
S=   \int  \rd^4x \rd^4 \q  \rd^4 \bar \q \, E\,{\mathfrak K}(X^I, \bar X^{\bar J} )~, \qquad
{\bar \cD}^{\ad }_{\hat i} X^I=0~,
\eea
where $\mathfrak K$ is the K\"ahler potential of ${\mathfrak M}^n$. The induced action
\eqref{6.16} is a special member of this family.
\\

\noindent
{\bf Acknowledgements:}\\
We are grateful to Joshua Pinelli for assistance with a calculation.
The work of SK is supported in part by the Australian Research Council, project No. DP200101944.


\appendix

\section{Proofs} \label{appA}

In this appendix we provide the essential steps in the proofs of the two important identities \eqref{id} and \eqref{mr}.

\subsection{Identity \eqref{id}} \label{proof1}
Here we establish the identity
\begin{align}
\int\! \dd k\, (k^2)^{j-1} H_j [B_1 \otimes B_2 \otimes \cdots & \otimes B_j] =(-\ii)^{j-1} h\,  M^{-1} B_1 M^{-1} B_2 M^{-1} B_3 \cdots M^{-1} B_j M^{-1}  \label{p1}
\end{align}
with $j\geq 1$, for arbitrary operators $B_i$.

We begin by generating an expression for the right-hand-side of \eqref{p1} by repeatedly employing the following identity, which is equation \eqref{PMid}:
\begin{align}
P M^{-m}= \sum_{n=0}^{\infty}\frac{(n+m-1)!}{n!(m-1)!} M^{-(n+m)} L_M^n (P)   \label{PMid2}
\end{align}
for any operator $P$.   We use the above identity -- working from right to left -- to pull all factors of negative powers of $M$ to the left of the expression on the right-hand-side of \eqref{p1}.  The result is:
\begin{align}
&(-\ii)^{j-1} h\,  M^{-1} B_1 M^{-1} B_2 M^{-1} B_3 \cdots M^{-1} B_j M^{-1} \no \\ &=(-\ii)^{j-1} h\!\!\!\!\! \sum_{n_1,n_2,\ldots,n_j =0}^{\infty} \frac{(n_1+n_2+\cdots +n_j+j)! }{n_1!n_2! \cdots n_j!(n_j+1)(n_j+n_{j-1}+2)\cdots (n_j+n_{j-1}+\cdots+n_1 + j)}  \no \\ &\qquad\qquad\qquad\qquad \times M^{-(n_1+n_2+ \cdots n_j+j+1)}L_M^{n_1} (B_1)L_M^{n_2} (B_2) \cdots L_M^{n_j} (B_j)\,. \label{p2}
\end{align}
We now show that this is equivalent to the left-hand-side of \eqref{p1} by directly using the definition of $H_n$ (expression \eqref{defHn}) as follows:
\begin{align}
&\int\! \dd k\, (k^2)^{j-1} H_j [B_1 \otimes B_2 \otimes \cdots  \otimes B_j] \no \\&= \int\! \dd k\, (k^2)^{j-1} \ex^A \int_{0}^{1} \! dy_1\int_{0}^{y_1} \! dy_2 \ldots \int_{0}^{y_{j-1}}\! dy_{j} \left(\ex^{-y_1 A}B_1\ex^{y_1 A}\right) \times \ldots \times \left(\ex^{-y_j A}B_j\ex^{y_j A}\right)\no
\\&= \int\! \dd k\, (k^2)^{j-1} \ex^A \int_{0}^{1} \! dy_1\int_{0}^{y_1} \! dy_2 \ldots \int_{0}^{y_{j-1}}\! dy_{j}  \no\\
& \qquad\qquad\qquad \sum_{n_1,n_2,\ldots,n_j =0}^{\infty} \left(\frac{(- y_1)^{n_1}}{n_1!}L_A^{n_1} (B_1)\right)\left(\frac{(- y_2)^{n_2}}{n_2!}L_A^{n_2} (B_2)\right) \cdots \left(\frac{(- y_j)^{n_j}}{n_j!}L_A^{n_j} (B_j) \right)\no
\\&= \int\! \dd k\, (k^2)^{j-1} \ex^A \sum_{n_1,n_2,\ldots,n_j =0}^{\infty} \frac{(- 1)^{n_1+n_2+\cdots +n_j }}{n_1!n_2! \cdots n_j!}\no \\ &\qquad\qquad \times \frac{L_A^{n_1} (B_1)L_A^{n_2} (B_2) \cdots L_A^{n_j} (B_j)}{(n_j+1)(n_j+n_{j-1}+2)(n_j+n_{j-1}+n_{j-2}+3) \cdots (n_j+n_{j-1}+\cdots+n_1 + j)}\no\\&= \sum_{n_1,n_2,\ldots,n_j =0}^{\infty}   \frac{(- 1)^{n_1+n_2+\cdots +n_j }}{n_1!n_2! \cdots n_j!} \frac{(-i)^{n_1+n_2+\cdots +n_j}} {(n_j+1)(n_j+n_{j-1}+2)\cdots (n_j+n_{j-1}+\cdots+n_1 + j)}
\no \\& \qquad \qquad\qquad  \times \left(\int\! \dd k\, (k^2)^{n_1+n_2+ \cdots n_j + j-1} \ex^{- \ii k^2 M}  \right)L_M^{n_1} (B_1)L_M^{n_2} (B_2)\cdots L_M^{n_j} (B_j)
\no \\ & =(-\ii)^{j-1} h\!\!\!\!\! \sum_{n_1,n_2,\ldots,n_j =0}^{\infty} \frac{(n_1+n_2+\cdots +n_j+j)! }{n_1!n_2! \cdots n_j!(n_j+1)(n_j+n_{j-1}+2)\cdots (n_j+n_{j-1}+\cdots+n_1 + j)}  \no \\ &\qquad\qquad\qquad\qquad \times M^{-(n_1+n_2+ \cdots n_j+j+1)}L_M^{n_1} (B_1)L_M^{n_2} (B_2) \cdots L_M^{n_j} (B_j)
\end{align}
where the last expression is \eqref{p2}, the right-hand-side of \eqref{p1}\,.

\subsection{Identity \eqref{mr}} \label{proof2}
In this subsection we use some of the results from \cite{Iochum:2016ynh} to sketch the details of an alternative proof of identity \eqref{mr}.  For the purposes of clarity, all summations in this subsection will be made explicit.

First we are interested in finding an expression for integrals of the form
\begin{align}
\tr \int\! \dd k \, (k^2)^{(n-2)} H_n [B_1 \otimes \cdots \otimes B_n] \qquad \qquad n \geq 2\,, \label{appeq}
\end{align}
where the $B$'s are all matrices (i.e. not differential operators).   Introducing a spectral decomposition for positive definite matrices $M$, we write
\begin{align}
M = \sum_i \l_i E_i\,,
\end{align}
where the sum is over all eigenvalues $\l_i$, and the projection matrices $E_i$ satisfy $E_i E_j = \d_{ij}E_i$.  We also introduce the following notation
\begin{align}
 \int\! \dd y  =  \int_{0}^{1} \! \dd y_1\int_{0}^{y_1} \! \dd y_2 \ldots \int_{0}^{y_{n-1}}\! \dd y_{n}\,.
\end{align}
Noting the definition of $H_n$, equation \eqref{defHn}, and after performing the $k$-integral, expression \eqref{appeq} becomes:
\begin{align}
\tr & \int\! \dd k \, (k^2)^{(n-2)} H_n [B_1 \otimes \cdots \otimes B_n]  \no \\ &=
 (-\ii)^{n-2} (n-1)! h\!\!\! \sum_{i_0,i_1,i_2,\ldots,i_n}\int\! \dd y\no
\Big(\l_{i_0}+ \left(\l_{i_1}-\l_{i_0}\right)y_1 + \cdots + \left(\l_{i_n}-\l_{i_{n-1}}\right)y_n\Big)^{-n}  \no \\&
\qquad\qquad \times\tr\Big(E_{i_0} B_1 E_{i_1} B_2 E_{i_2} \ldots B_n E_{i_n}\Big)\,.
\end{align}
Following \cite{Iochum:2016ynh} we define the continuous functions
\begin{align}
I_{\a,n}(\l_0,\l_1, \ldots \l_n):=\int\! \dd y\Big(\l_{0}+ \left(\l_{1}-\l_{0}\right)y_1 + \cdots + \left(\l_{n}-\l_{{n-1}}\right)y_n\Big)^{-\a}\,,
\end{align}
for which we are interested in $\a=n$.  It was demonstrated in \cite{Iochum:2016ynh} that for $n\geq1$
\begin{align}
I_{n,n}(\l_0,\l_1, \ldots \l_n) = \frac{(-1)^{n-1}}{(n-1)!} \sum_{i=0}^{n}\Big[\prod_{\substack{j=0 \\j\neq i}}^n(\l_i - \l_j)^{-1}\Big] \ln \l_i\,.
\end{align}
Noting that
\begin{align}
  \tr\Big(E_{i_0} B_1 E_{i_1} B_2 E_{i_2} \ldots B_n E_{i_n}\Big) = \d_{i_0,i_n}\tr\Big( B_1 E_{i_1} B_2 E_{i_2} \ldots B_n E_{i_n}\Big)\,,
\end{align}
we write
\begin{align}
J_n(\l_1, \ldots \l_n) := \lim_{\l_0\rightarrow\l_n}  I_{n,n}(\l_0,\l_1, \ldots \l_n)\,,
\end{align}
which is well-defined, followed by
\begin{align}
F_n(\l_1, \ldots \l_n) := J_n(\l_1, \ldots \l_n) + \textrm{cyclic perms of $\l_i$}\,.
\end{align}
It now follows that our integrals involving the cyclic combinations $C_n$, defined in \eqref{defcn}, can be expressed as
\begin{multline}
\tr \int\! \dd k \, (k^2)^{(n-2)} C_n [B_1 \otimes \cdots \otimes B_n] \\
= (-\ii)^{n-2} (n-1)! h\!\!\! \sum_{i_1,i_2,\ldots,i_n}F_n(\l_{i_1}, \ldots \l_{i_n}) \tr\Big(B_1 E_{i_1} B_2 E_{i_2} \ldots B_n E_{i_n}\Big)\,.
\end{multline}
With some work it can be shown that all of the logarithm contributions of the functions $F_n$ cancel, and ultimately they reduce to
\begin{align}\
F_n(\l_1, \ldots \l_n) = \frac{1}{(n-1)!}\frac{1}{\l_1 \l_2\cdots \l_n}\,.
\end{align}
It then follows that
\begin{align}
\tr & \int\! \dd k \, (k^2)^{(n-2)} C_n [B_1 \otimes \cdots \otimes B_n]  =
 (-\ii)^{n-2}  h\, \tr\Big(B_1 M^{-1} B_2 M^{-1} \ldots B_n M^{-1}\Big)\,,
\end{align}
which is identity \eqref{mr}.


\section{Invariance of $a_2$ under local operator rescalings} \label{proof3}

Consider a second-order operator $\D$  acting on a space of fields $\vf^i$, and let
$a_2 $ be the corresponding $a_2$ coefficient.
In this appendix we demonstrate that
\bea
{\rm Tr} \, a_2 := \int \rd^4x \, \sqrt{-g}\,  {\rm tr} \Big(a_2(x,x) \Big)
\eea
is invariant under local transformations
\bea
\D ~\to ~ \tilde \D = M(x) \D~,
\eea
where $M(x)$ is a non-singular $x$-dependent matrix. The arguments below are quite general and are valid in curved space.

Starting from the formal relations
\bea
\re^{\ri \,\G} = {\rm Det}^{-\hf} \D~, \qquad \re^{\ri \,\tilde \G}
= {\rm Det}^{-\hf} \tilde \D = {\rm Det}^{-\hf} \big( M \D \big)~,
\eea
we introduce the regularised effective actions
\bea
\G_\o = \frac{\m^{2 \w}}{2}\int_{0}^{\infty} \frac{\dd s}{ (\ii s)^{1-\w}}
{\rm Tr} \,\big( \re^{\ri s\D} \big)~,\qquad
\tilde{\G}_\o = \frac{\m^{2 \w}}{2}\int_{0}^{\infty} \frac{\dd s}{ (\ii s)^{1-\w}}
{\rm Tr} \,\big( \re^{\ri s \tilde \D} \big)~.
\eea
In the infinitesimal case, $M = {\mathbbm 1} + \d M$, for the variation of the effective action we get
\bea
\d_M \G_\o =  \frac{\m^{2 \w} }{2}\int_{0}^{\infty} \rd s\, (\ri s)^{\w}
{\rm Tr} \,\Big( \d M \D \re^{\ri s\D} \Big)
= -  \frac{\m^{2 \w}}{2} \o\int_{0}^{\infty} \frac{\dd s}{ (\ii s)^{1-\w}}
{\rm Tr} \,\Big(\d M \re^{\ri s\D} \Big)~.
\eea
In the limit $\o \to 0$, we obtain
\bea
\lim_{\o\to 0} \d_M \G_\o = -\frac{1}{2(4\pi)^2} {\rm Tr} \Big( \d M a_2\Big)~.
\eea
We see that the operator deformation
$\D ~\to ~ \tilde \D =  \D + \d M(x) \D$ is accompanied by
 a finite local variation of the effective action. The logarithmically divergent part of the effective action remains unchanged.

\section{The coefficient  $\bm{a}_2^c$} \label{appB}

Here we display a final expression for the chiral coefficient $\bm{a}_2^c$, the result of the calculation in section \ref{sec:chiral}.  For ease of computational reproducibility we have not used any of the chirality constraints \eqref{cc} in simplifying this result:
\begin{multline}
\bm{a}_2^c = \tr\Bigg\{-4\~G \~N-\frac12 \~G\~L^{a\a}\~L_{a\a}+ \~F^\a\~F_\a -\frac12\~J_a \~L^{a \a} \~F_\a -\frac12 \~L^{a \a}\~J_a \~F_\a+ \~N \~J^a\~J_a \\+\frac{1}{24}\Big(2\~J^a\~J_a\~L^{b\a}\~L_{b\a}
+2\~J_a\~J^b\~L^{a\a}\~L_{b\a}
+2\~J^a\~J_b\~L^{b\a}\~L_{a\a}
+\~J_a\~L^{a\a}\~J^b\~L_{b\a}
+\~J^a\~L^{b\a}\~J_a\~L_{b\a}
+\~J^a\~L^{b\a}\~J_b\~L_{a\a}\Big)\\-\~F^{\a}(\pd^a \~L_{a\a})-4\~F^{\a}(D_\a \~N)
+\frac12 \~F^{\a}\~L_{a\a}(D^\b \~L^a_{\phantom{a}\b} )
-\frac12 \~F^{\a}(D_\a \~L^{a\b} )\~L_{a\b}
-2\~J^a (\pd_a \~N)
+ \~J_a \~L^{a\a}(D_\a \~N)\\
- \~J^a\~N (D^\a \~L_{a\a})
+\frac{1}{6}\Big(
-\~J^a \~L^{b\a}(\pd_a\~L_{b\a})
-2 \~J^a (\pd_a\~L^{b\a})\~L_{b\a}
-\~J^a \~L^{b\a}(\pd_b\~L_{a\a})
+2\~J^a \~L_{a}^{\phantom{a}{\a}}(\pd^b\~L_{b\a})\\
+\~J^a (\pd_b\~L^{b\a}) \~L_{a\a}
+\~J_a (\pd^b\~L^{a\a})\~L_{b\a}\Big)
+\frac{1}{12}\Big(-\~J_a \~L^{a\a}\~L_{b\a}(D^\b L^{b}_{\ph{b}\b})
+\~J_a \~L^{a\a}(D_\a \~L^{b\b}) \~L_{b\b}\\
+\~J^a \~L^{b\a}(D_\a \~L_b^{\ph{b}\b}) \~L_{a\b}
-\~J^a \~L^{b\a}\~L_{a\a}(D^\b \~L_{b\b})
-\~J^a \~L^{b\a}\~L_{b\a}(D^\b \~L_{a\b})
+\~J_a \~L^{b\a}(D_\a \~L^{a\b}) \~L_{b\b}\Big)\\
+2 (D^\a\~ N)(D_\a\~ N)
+\frac{1}{12}\Big(2(\pd_a \~L^{a\a})(\pd^b \~L_{b\a})+(\pd^a \~L^{b\a})(\pd_a \~L_{b\a})\Big)
+\frac{2}{3}\Big((\pd^a \~N)(D^\a \~L_{a\a})\\
+2(\pd_a \~L^{a\a})(D_\a \~N)\Big)
-\frac12(D^\a \~L^{a\b})D_\a(\~L_{a\b}\~N)+\~N (D^\a \~L^{a}_{\ph{a}\a})D^\b(\~L_{a\b})\\
+\frac{1}{36}\Big(-7 (\pd_a \~L^{a\a})\~L^{b\b}(D_\a \~L_{b\b})
+7 (\pd_a \~L^{a\a})(D_\a \~L^{b\b})\~L_{b\b}
-6 (\pd_a \~L^{b\a})(D_\a \~L^{a\b})\~L_{b\b}\\
-2 (\pd^a \~L^{b\a})(D_\a \~L_{b}^{\ph{b}\b})\~L_{a\b}
+10 (\pd_a \~L^{a\a})\~L^{b\b}(D_\b\~L_{b\a})
-4 (\pd^a \~L^{b\a})\~L_{a\a}(D^\b\~L_{b\b})\\
+5 (\pd^a \~L^{b\a})\~L_{b\a}(D^\b\~L_{a\b})
-7 (\pd^a \~L^{b\a})(D^\b\~L_{a\b})\~L_{b\a}
+10 (\pd^a \~L^{b\a})(D^\b\~L_{a\a})\~L_{b\b}\Big)\\
+\frac{1}{96}\Big(
\~L^{a\a} \~L_{a\a}(D^\b\~L^b_{\ph{b}\b})(D^\g\~L_{b\g})
-\~L^{a\a} \~L_{a\a}(D^\b \~L^{b\g})(D_\g \~L_{b\b})
+\~L^{a\a} \~L^b_{\ph{b}\a}(D^\b \~L_{a\b})(D^\g \~L_{b\g})\\
+\~L^{a\a} \~L_{b \a}(D^\b \~L^b_{\ph{b}\b})(D^\g \~L_{a\g})
-\~L^{a\a} \~L^b_{\ph{b} \a}(D^\b \~L_a^{\ph{a}\g})(D_\g \~L_{b\b})
-\~L^{a\a} \~L_{b \a}(D^\b \~L^{b\g})(D_\g \~L_{a\b})\\
+\~L^{a\a} \~L_{a}^{\ph{a}\b}(D_\b \~L^{b\g})(D_\g \~L_{b\a})
+\~L^{a\a} \~L^{b \b}(D_\b \~L_a^{\ph{a}\g})(D_\g \~L_{b\a})
+\~L^{a\a} \~L^{b \b}(D_\b \~L_b^{\ph{b}\g})(D_\g \~L_{a\a})\\
-\~L^{a\a} \~L_a^{\ph{a} \g}(D^\b \~L^b_{\ph{b}\b})(D_\g \~L_{b\a})
-\~L^{a\a} \~L^{b \g}(D^\b \~L_{a\b})(D_\g \~L_{b\a})
-\~L^{a\a} \~L^{b \g}(D^\b \~L_{b\b})(D_\g \~L_{a\a})\\
-2\~L^{a\a} (D_\a \~L_a^{\ph{a}\b})\~L^b_{\ph{b}\b}(D^\g\~L_{b\g})
-2\~L^{a\a} (D_\a \~L^{b\b})\~L_{a\b}(D^\g\~L_{b\g})
-2\~L^{a\a} (D_\a \~L^{b\b})\~L_{b\b}(D^\g\~L_{a\g})\\
+\~L^{a\a} (D^\g \~L_a^{\ph{a}\b})\~L^b_{\ph{b}\b}(D_\g\~L_{b\a})
+\~L^{a\a} (D^\g \~L^{b\b})\~L_{a\b}(D_\g\~L_{b\a})
+\~L^{a\a} (D^\g \~L^{b\b})\~L_{b\b}(D_\g\~L_{a\a})
\Big)
\Bigg\}\,. \label{ca2appB}
\end{multline}
It can be shown by using the  chirality constraints \eqref{cc} that $\bm{a}_2^c$ is chiral.


\begin{footnotesize}

\end{footnotesize}

\end{document}